\documentstyle [11pt] {article}

\makeatletter
\def\appsection{\@startsection {section}{1}{\z@}{-3.5ex plus -1ex minus
 -.2ex}{2.3ex plus .2ex}{\Large\bf \noindent Appendix }}
\makeatother

{\catcode `\@=11 \global\let\AddToReset=\@addtoreset}
\AddToReset{equation}{section}

\newcommand{\N}{I\!\! N}
\newcommand{\R}{I\!\! R}
\newcommand{\Z}{Z\!\!\!\!Z}

\newcommand{\scri}{{\cal I}} 
\def\SCRI{Scri}
\def\Scri{Scri}

\def\cosec{\mathop{\rm cosec}\nolimits}

\newcommand{\lint}{loc}
\newcommand{\C}[1]{\mbox{$C_\infty^{#1}$}}
\newcommand{\A}{\mbox{${\cal A}^{phg}$}}

\newcommand{\commentout}[1]{}

\newcommand{\cmm}[1]{}

\newcommand{\lhs}{\mbox{\{l.h.s.\}}}
\newcommand{\rhs}{\mbox{\{r.h.s.\}}}
\newcommand{\eq}[1]{(\ref{#1})}
\newcommand{\beaa}{\begin{eqnarray*}}
\newcommand{\bea}{\begin{eqnarray}}
\newcommand{\be}{\begin{equation}}
\newcommand{\bnh}{\bar{\cal N} }

\newcommand{\bu}{\breve u}
\newcommand{\bau}{\bar u}

\newcommand{\eeaa}{\end{eqnarray*}}
\newcommand{\eea}{\end{eqnarray}}
\newcommand{\ee}{\end{equation}}

\newcommand{\is}{\Sigma}

\newcommand{\mt}{{}^2\!{\cal M}}
\newcommand{\nh}{{\cal N} }
\newcommand{\nhb}{\bar{\cal N} }
\newcommand{\nn}{\nonumber}

\newcommand{\ra}{\rightarrow}
\newcommand{\rai}{\rightarrow \infty}
\newcommand{\raz}{\rightarrow 0}
\newcommand{\spm}{\gamma }
\newcommand{\spt}{{\cal M} }
\newcommand{\sts}{space--times}
\newcommand{\st}{space--time}

\newcommand{\tr}{{\rm tr}\,}
\renewcommand{\pt}{\partial}

\newtheorem{Proposition}{Proposition}

\AddToReset{Conjecture}{section}
\AddToReset{Definition}{section}
\AddToReset{Lemma}{section}
\AddToReset{Corollary}{section}
\AddToReset{Proposition}{section}
\AddToReset{Theorem}{section}

\begin{document}

\title{
Gravitational waves in general relativity: XIV\thanks{Part XIII of
this series (by H.\ Bondi and F.A.E.\ Pirani) appeared in Proc.\ Roy.\
Soc.\ {\bf A421}, 395 (1989)}. Bondi expansions and the
``polyhomogeneity'' of \SCRI}

\author{Piotr T.\ Chru\'sciel\thanks{Alexander von Humboldt fellow.
On leave from the Institute of
Mathematics, Polish Academy of Sciences, Warsaw.
Supported in part by
the KBN grant \# 2 1047 91
01.
E--mail:
piotr@ibm-1.mpa.ipp-garching.mpg.de} \\Max Planck
Institut f\"ur Astrophysik\\ Karl Schwarzschild Strasse 1\\ D--8046
Garching bei M\"unchen, Germany\\
\\
Malcolm A.H.\ MacCallum\thanks{E--mail: M.A.H.MacCallum@qmw.ac.uk}
\\School of Mathematical Sciences \\
Queen Mary and Westfield College, University of London \\
Mile End Road, London E1 4NS, U.K.\\
\\
David B.\ Singleton\thanks{E--mail: D.Singleton@anu.edu.au} \\ Centre
for Scientific Computing
 \\ Australian National University\\
 Canberra A.C.T. 2601, Australia
 \\}

\maketitle

\begin{abstract}
The structure of {\em polyhomogeneous} \sts\ ({\em i.e.}, \sts\ with
metrics which admit an expansion in terms of $r^{-j}\log^i r$)
constructed by a Bondi--Sachs type
method is analysed. The occurrence of some $\log$ terms in an asymptotic
expansion of the metric is related to the non--vanishing of the Weyl tensor
at \Scri. Various quantities of interest, including the Bondi mass loss
formula, the peeling--off of the Riemann tensor and the Newman--Penrose
constants of motion are re-examined in this context.
\end{abstract}

\section{Introduction}
\label{introduction}

In general relativity an important question
is: what does the gravitational field of a radiating
asymptotically Minkowskian system look like?
The answer to that question proposed by Bondi {\em et
al.\ }\cite{BMS}, Sachs \cite{Sachs} and Penrose \cite{Penrose} seems
to have been
adopted by researchers ({\em cf.\ e.g.\
}\cite{Wald,NewmanTod}), in spite of the wide evidence against
this proposal: indeed it has been suggested both by the analysis of
Christodoulou and Klainerman \cite{ChKl}   and by various approximate
calculations ({\em cf.\ e.g.\
}\cite{Damour} and references therein) that such systems  generically do
not satisfy the Bondi--Penrose--Sachs asymptotic conditions.
In a recent study \cite{ACh} ({\em cf.\ }also \cite{ACF}) of the asymptotic
properties of solutions of constraint equations on spacelike hypersurfaces
intersecting ``\Scri'' transversally it has similarly been
observed that generic Cauchy data constructed in such a setting by the
``conformal method'' failed to be smoothly extendable, after appropriate
rescalings, to the conformal boundary.
More precisely, it has been shown ({\em cf.\ }\cite{ACh,ACWeyl} for
more details) that, when considering Cauchy data constructed by the conformal
method with smooth up to boundary ``seed'' fields and with the condition $\tr
K={\rm const}\ne 0$, one has:
\begin{enumerate}
\item
generically, for such data no conformal factor $\Omega$ exists for which the
shear of $\scri$ in the metric $\Omega^2\gamma$ vanishes at $\pt\is$; by
the vanishing of the shear of $\scri$ we mean the somewhat stronger
statement
that
\begin{equation}
\nabla_\mu\nabla_\nu\,\Omega\Big|_{\scri}=0 \label{shear}
\end{equation}
(recall that in the case of $C^2(\bar\spt)$ metrics the existence of an
$\Omega$ such that (\ref{shear}) holds
follows from the vacuum field equations, {\em cf.\ e.g.\ }\cite{Wald});
\item
consider those data for which the shear of $\scri$ vanishes for
appropriately chosen $\Omega$. Generically, for such data the Weyl tensor of
$\Omega^2\gamma$ {\em does not} vanish at $\scri$.
(Recall that for vacuum metrics $\gamma$ such that $\Omega^2\gamma$
is $C^3$ up to boundary on $\tilde \spt$ the vanishing of the Weyl tensor
of $\Omega^2\gamma$ at $\scri$ follows by a theorem of Penrose
\cite{Penrose,PenroseRindler,Geroch EW}).
\end{enumerate}

The results obtained in \cite{ACh} seem to indicate very strongly that
a consistent
setup in which the gravitational radiation field can be described
in {\em generic }situations
is that of manifolds $(\tilde\spt,\tilde\gamma)$, $\tilde \gamma
=\Omega^2\gamma$ with metrics $\tilde \gamma$ which are not smooth
but {\em polyhomogeneous}\footnote{The term {\em polyhomogeneous}
seems to have been adopted in the mathematical literature for the
kind of expansion considered here,
{\em cf.\ e.g.\ }\cite{Mazzeo}. Gelfand and Shilov
\cite{GS} use the term {\em associated homogeneous} for a similar
notion. The members of the Garching
relativity seminar have suggested use of the term {\em polylogarithmic}
for this kind
of expansion. Winicour \cite{Winicour} uses the term {\em logarithmic
asymptotic flatness} in a somewhat similar setting.} near $\scri$.
(A function $f$ is
called {\em polyhomogeneous} if it admits an expansion in terms of
$r^{-j}\log^ir$ rather than $r^{-j}$, {\em cf.\ }Appendix \ref{conventions}
for a more precise definition.)
The object of this paper is to show that at least part of the results
described above can be obtained in a rather simpler way in a
Bondi--Sachs type setting, as set out in earlier papers in this series
(Papers VII, VIII and IX, referred to here, frequently, as \cite{BMS},
\cite{Sachs}, and \cite{vdB} respectively).

In section 2,
we show that the hypothesis of polyhomogeneity of $\scri$ is {\em
formally consistent\/} with the Einstein equations.
(Note, however, that thanks to the important theorems of Friedrich
\cite{Friedrich1,Friedrich2,FrJDG}, together with the results of
\cite{ACh,ACF}, a large class of \sts\ satisfying the Bondi--Penrose--Sachs
conditions is now known to exist. On the other hand no proof that the
Cauchy problem is well posed for polyhomogeneous but not smooth initial
data of ``hyperboloidal'' type is available yet.)\
We show that the characteristic initial value problem of Bondi--Sachs type
is formally well posed in the space of polyhomogeneous metrics, in the
sense that the
(retarded)
time derivatives of the fields on the initial data hypersurface are
polyhomogeneous if the free initial data are (with the same ``degrees of
polyhomogeneity''
when these degrees are chosen appropriately, {\em cf.\/}
Section \ref{analysis} for details),
and, in a manner
completely analogous to that of the original Bondi--Sachs analysis,
that one can
write down a hierarchy of evolution equations for the coefficients of the
polyhomogeneous expansion of the free data.

In section 3 we show that
in the class of \sts\
considered in this paper the conformal factor
$\Omega$ can
always be chosen so that (\ref{shear}) holds. Thus, Cauchy data
incompatible with (\ref{shear}) cannot lead to a \st\ of the type considered
here ({\em cf.\ }\cite{ACWeyl} for a similar result in a somewhat
different setting).
We prove that initial data,
constructed by a Bondi--Sachs procedure starting from free data smooth at
$\scri$, will be smooth at $\scri$ if and only if the free initial
data are such
that the Weyl tensor of $\Omega^2\gamma$ vanishes at $\scri$.
We find that the
Trautman--Bondi mass loss formula \cite{Trautman,BMS,Sachs}
remains unchanged in
the polyhomogeneous case; thus the  Bondi mass
 is still well defined, and is
a monotonically decreasing function of retarded time\footnote{More
precisely,
for all polyhomogeneous metrics there is a quantity which
we call the Bondi mass, which  is a nonincreasing function of retarded
time, and which reduces to the quantity defined by Bondi when Bondi's
hypotheses are satisfied.  We believe that the ``real mass'' should
not be
defined {\em ad hoc}, but
by a limiting procedure involving perhaps the Freud ``superpotential'' for
Einstein's energy, as done {\em e.g.\ }by Trautman in \cite{Trautman}.
If one does
that, we expect that one will find equality of the quantity we
define as the Bondi mass with the quantity obtained from the limiting
procedure only when $V$ has {\em no }$\log^N$ terms, {\em i.e.}, when the
logarithmic terms in $V$ start at the $r^{-i}$ level, with some $i\ge 1$.
A precise formulation of such statements lies
outside the scope of this paper.}.
We also note
that for a class of
polyhomogeneous metrics the peeling--off property of the Riemann
tensor is the same as the one for smooth
metrics up to
$O(r^{-2-\epsilon})$, $0\le\epsilon<1$, terms.
We show that some quantities
 built out of the restriction of the Weyl tensor to $\scri$ are
(in general nontrivial) constants
of motion, as already noted by Winicour \cite{Winicour} and by
Christodoulou and
Klainerman \cite{ChKl}. More generally, we find (in section 2) that the
``leading $\log$
coefficients'' of the polyhomogeneous expansion are constants of motion.
We argue, from  an explicit calculation in the axisymmetric case,
that the Newman--Penrose
constants of motion \cite{NPCM,PenroseRindler} cease to be constants of
motion in generic polyhomogeneous situations,  although our
example does give a new constant of the motion. (It could,
therefore, be
that some new functionals of the field, which reduce to the
Newman--Penrose constants of motion when $\scri$ is smooth, are
constants of motion in the polyhomogeneous situation. We do not have an
answer to that question.) Section 4 considers the construction of Bondi
coordinates in our more general setting.

The results of our analysis show that the presence of some $\log r$
terms in an
asymptotic expansion of the metric is quite natural, and does not lead to
any serious extra difficulties in the analysis of the geometry. Recall
that the
imposition of the conditions which lead to the vanishing of the $\log$ terms
was interpreted in some earlier papers of this series as an {\em outgoing
radiation condition\/} \cite{BMS,Sachs,vdB}. Two concerns had to be
addressed: the possibility of advanced rather than retarded solutions,
and the possibility of retarded waves travelling in the inward radial
direction but at indefinitely large distances. With the help of our
present understanding of \Scri, it is clear that if \Scri$^+$ is
well-defined, as it is here, there is no advanced wave involved, and that
a \st\ has purely outgoing
radiation if and only if there is no radiation at \Scri$^-$
({\em cf.\ }also \cite{LeipoldWalker} for a similar point of view and
for explanation of the difficulties that arise with local
characterization of incoming and outgoing parts of the field even for
linear theories in flat space). Since the space-times discussed here
can satisfy both these requirements, we can safely abandon the
``outgoing radiation'' condition of \cite{BMS}.

Moreover, we note that
there exists a family of electrovacuum ``small data" \sts\ constructed by
Cutler and Wald \cite{CW}, and also a family of ``small data"
Einstein--Yang--Mills spherically symmetric \sts\ constructed by Bartnik
\cite{BartnikPerjes}, which have the following properties: they possess a
smooth \Scri$^+$ and a smooth \Scri$^-$, and decay to a smooth $i^+$ in
the future and a smooth $i^-$ in the past. Because the metric decays
smoothly both
in the future and the past there is both {\em outgoing\/} and {\em
incoming\/} radiation in those \sts. Since both \Scri's are smooth, the
``outgoing radiation condition" holds at \Scri$^+$, and an analogous
``incoming radiation condition" holds at \Scri$^-$, which is clearly absurd.
These examples show that not only is the vanishing of the $\log$ terms
at \Scri\ unnecessary to ensure ``outgoing radiation'', it also does
not prevent incoming radiation. We conclude that the association of
the absence of $\log r$ terms with ``outgoing radiation'' lacks
justification, and that a better formulation of an ``outgoing
radiation'' condition could probably be given in terms of constancy of
the Bondi mass at \Scri$^-$.

In the discussion above we have adopted what we consider to be now a
standard notion of
``incoming'' and ``outgoing'' radiation. In particular,
it is clear, from the hyperbolic
nature of the Einstein equations, that  whenever a conformal
completion of the space--time exists in which the conformal boundary
is an {\em incoming null topological surface}, then there can be {\em no
influx} of gravitational radiation (or, for that matter, of any non--tachyonic
matter fields) through the surface in question.
Thus, the existence of a
conformal completion
of the above described nature
guarantees that we have
an isolated system evolving in a self--consistent way, regardless of
whether or not the fields are asymptotically Minkowskian\footnote{As
pointed out below, several results proved in this paper (in particular the
self--consistency of the polyhomogeneous setup) will still be true if
the ``sphere of null directions'' $S^2$ is replaced by an arbitrary
two--dimensional, perhaps but not necessarily compact, manifold $M^2$.
Examples of vacuum space--times with such an asymptotic structure
(and actually a smooth \Scri) are given by {\em e.g.\/} some
Robinson--Trautman space--times.},
regardless of the decay rates of the fields towards the conformal
boundary, the degrees of differentiability of some
perhaps conformally rescaled fields at the
conformal boundary, etc.
In view of that observation it might not be so surprising that for
the { polyhomogeneous} \Scri's
considered here the Trautman--Bondi mass--loss law
holds, regardless of the occurrence of some perhaps high powers of
$\log r$ in the $1/r$ terms in the metric.

It should be pointed out that several of the results discussed here
have already been observed in a similar
setting by
Winicour \cite{Winicour}. (However, we learned about Ref.\
\cite{Winicour} only after most of  the work presented here was completed.
Also it seems that in \cite{Winicour} emphasis is put
on somewhat different
issues.)
In
this context one should also mention the results of Novak and
Goldberg \cite{NovakGoldberg} ({\em cf.\ }also
\cite{CouchTorrence,Moreschi}),
who perform a somewhat similar analysis of  the Newman--Penrose
equations on a null initial hypersurface.

\section{The Bondi--Sachs characteristic initial value problem}
\label{analysis}

In this section we shall consider the initial value problem for space-times
$(\spt, \spm)$ with a metric of the form
\be
\spm_{\mu\nu}\,dx^\mu\,dx^\nu = -\frac{V e^{2\beta}}{r}\, du^2 -
2e^{2\beta}\, du\,
dr
+ r^2\, h_{ab} (dx^a + U^a du) (dx^b + U^b du)\,.
\label{(B5.1)}
\ee
We shall mainly be interested in the behaviour of $\spm$ on
the hypersurface\footnote{Most of the analysis presented here goes through
when $S^2$ is replaced by any two dimensional, compact, orientable
manifold $\mt$.}
$$\nh = \{u = 0,\, r \ge R,\, x^a \in S^2\}\,,$$
where $S^2$ is topologically a two dimensional sphere.
 (The question of the existence of
coordinate systems in which an asymptotically Minkowskian metric takes
the form (\ref{(B5.1)}) is considered in Section \ref{coordinates}.)\ As has
been analysed by Bondi {\em et al.\/}\ \cite{BMS} in the axisymmetric
case and by Sachs \cite{Sachs} in general ({\em cf.\ }also \cite{vdB}),
to construct a vacuum metric of
the form (\ref{(B5.1)}) one has to prescribe on $\nh$ the family of metrics
$h(r) \equiv h_{ab}(r, x^a)\, dx^a\, dx^b$ on $S^2$ parametrized by
$r$,
the family of vector fields $U(r) \equiv U^a(r, x^a) \partial_a$ on
$S^2$ parametrized by
$r$,
and the scalar fields $V$ and $\beta$. These quantities are not freely
specifiable but have to satisfy constraint equations:
\begin{equation}
\forall \quad X \in T \nh \qquad \qquad R_{\mu\nu}\, k^\mu\, X^\nu = 0\,,
\label{(B5.2)}
\end{equation}
where $k^\mu$ is any null vector field tangent to $\nh$ ({\em e.g.},
$k^\mu \pt_\mu = \pt_r$). As has been emphasized in \cite{BMS} and
\cite{Sachs}, the equations (\ref{(B5.2)}) do not impose any restrictions on
$h_{ab}\, dx^a\, dx^b$, and in fact can be viewed as equations which
together with appropriate boundary conditions determine $V$, $\beta$ and
$U^a\, \pt_a$ given $h_{ab}\, dx^a\, dx^b$. In \cite{BMS,Sachs} it was
shown that if we assume $h_{ab} \in C^\infty(\bnh)$ and moreover
\begin{eqnarray}
h_{ab}(r, x^a) = \hat h_{ab}(x^a) + \frac{h_{ab}^1(x^a)}{r} +
\frac{a(x^a)\hat h_{ab}(x^a)}{r^2} +
O\left(\frac{1}{r^3}\right)\,,
\label{(B5.3)}
\end{eqnarray}
for some functions
$\hat h_{ab}(x^a),h_{ab}^1(x^a),a(x^a)$,
then we will obtain
$r^{-2}V, \beta, U^a, \frac{\pt h_{ab}}{\pt u} \in
C^\infty(\bnh)$.

In \cite{BMS,Sachs} the absence of trace--free
$r^{-2}$ terms in (\ref{(B5.3)}) was termed the ``outgoing wave condition".
This condition
was imposed
{\em a priori\/} in \cite{BMS,Sachs} because the occurrence of the
trace--free terms led to
$r^{-j}\, \log^i r$ terms in $U^a, V$ and subsequently in $\frac{\pt
h_{ab}}{\pt u}$ --- this in turn led to $r^{-j}\, \log^i r$ terms in
$h_{ab}$ at any later moment of time.
It is therefore clear that a correct
setup for analysing the characteristic initial value
problem for metrics of the form
(\ref{(B5.1)}) is that of metrics $h_{ab}\, dx^a\, dx^b$ which are {\em
polyhomogeneous\/} to start with ({\em i.e.}, admit an
asymptotic expansion in terms of $r^{-j}\log^i r$). It is the aim of
this paper to re-examine
both the constraint and the evolution equations for metrics of the form
(\ref{(B5.1)}) in a polyhomogeneous setup.

Before proceeding to a detailed analysis of the Einstein equations,
let us first
consider the question of the boundary conditions satisfied by the fields
under consideration. Let therefore a polyhomogeneous metric $h_{ab}\,
dx^a\, dx^b$ (see Appendix \ref{conventions} for precise
definitions) be given, and suppose moreover that $h_{ab} \in C^0(\bnh)$
(if we write
$h_{ab} \in {\cal A}^{\{N_i\}}$, then the hypothesis $h_{ab} \in
C^0(\bnh)$ is equivalent to the condition $N_0 = 0$). It follows that the
limits
$$
\hat h_{ab} = \lim_{r\rai}\; h_{ab}
$$
exist, with $\hat h_{ab} \in C^\infty(S^2)$. As is well known,
({\em cf.\ e.g.\ }\cite{ChKl} for a simple and elegant proof)
 there exists a
diffeomorphism $\Phi : S^2 \ra S^2$ such that we have $\Phi^*\, {\hat
h} = \phi^2\, {\breve h}$, where $0 < \phi \in C^\infty(S^2)$ and ${\breve
h}$ is the standard round metric on $S^2$,
\be
\label{roundmetric}
{\breve h}_{ab}\, dx^a\, dx^b = d\theta^2 + \sin^2 \theta\,
d\varphi^2\,.
\ee
Replacing $(r, x^a)$ by $({\bar r}, {\bar x}^a) = (\phi r, \Phi^a(x^b))$ one
obtains a metric of the form (\ref{(B5.1)}) in which (dropping bars on
$\bar r,
{\bar x}^a$)
\begin{eqnarray}
\lim_{r\rai}\; h_{ab}\, dx^a\, dx^b = d\theta^2 + \sin^2 \theta\, d\varphi^2\,.
\label{(B5.4.0)}
\end{eqnarray}
It is not too difficult to show from eqs.\ (\ref{(B5.2)}) (which are
written out in detail in Appendix \ref{vdbeqs}) that under the condition
$h_{ab} \in {\cal A}_{phg}$ the limits
\begin{eqnarray}
H &\equiv &\lim_{r\rai}\; \beta\,,\label{(B5.4)}\\
X^a &\equiv & -\lim_{r\rai}\; U^a\label{(B5.5)}
\end{eqnarray}
exist, with $H, X^a \in C^\infty(S^2)$. Suppose for a moment that there
actually exists a \st\ with a metric of the form (\ref{(B5.1)}) on a set ${
U}_\epsilon \equiv \{u \in (-\epsilon,\, \epsilon),\, r > R,\, x^a \in
S^2\}$ with
some $\epsilon > 0$. (We should stress that in all the results obtained in
this section the hypothesis of the existence of an evolution of the initial
data defined on $ { U}_\epsilon$ for some $\epsilon > 0$ is {\em not
necessary\/}. This is due to the fact that all our analysis involves only
equations on $\nh$). Let $\psi(u, x^a)$ be the one parameter family of
diffeomorphisms of $S^2$ generated by the vector field $X = X^a\, \pt_a$;
we thus have
$$
\psi^a(u, x^b)\big|_{u=0} = x^a\,,
$$
$$
\frac{\pt\psi^a(u, x^b)}{\pt u} = X^a\big(u, \psi^c(u, x^b)\big)\,.
$$
It is easily seen that the coordinate transformation $(u, r, x^a) \ra
({\bar u},
{\bar r}, {\bar x}^a) = (u, r, {\bar x}^a)$, with ${\bar x}^a$ implicitly
defined by $x^a = \psi^a(u, {\bar x}^b)$, leads to a metric of the form
(\ref{(B5.1)}) for which we have
\begin{equation}
\lim_{r\rai}\; U^a = 0\,.
\label{(B5.6)}
\end{equation}
The above argument shows that the vector field $X^a\, \pt_a$ defined by
(\ref{(B5.5)}) has a gauge character, at least from a four dimensional point
of view. It should also be pointed out that the transformations leading to
(\ref{(B5.4.0)}) and (\ref{(B5.6)})
are compatible with an initial value
setup, because they do not deform the initial hypersurface $\nh$ in
\st. This shows that there is no loss of generality in assuming that
(\ref{(B5.4.0)}) and (\ref{(B5.6)}) hold\footnote{More precisely,
there is no loss of generality in assuming that (\ref{(B5.4.0)}) holds,
and  there is no loss of generality in assuming that
(\ref{(B5.6)}) holds
provided  the time--development of the data  is sufficiently regular
asymptotically in some neighbourhood of $\nh$ ({\em e.g.},
polyhomogeneous)
to be able to perform
the construction which leads to (\ref{(B5.6)}).}.

It is natural to ask whether $H$ defined by (\ref{(B5.4)}) can be
removed
by an appropriate choice of gauge. As shown by Bondi {\em et al.\/}
\cite{BMS} in the smooth axisymmetric case, and as shown in Section
\ref{coordinates} in the general polyhomogeneous case, the condition
\begin{equation}
\lim_{r\rai}\; \beta = 0
\label{(B5.7)}
\end{equation}
can always be achieved, at the price, however, of {\em deforming\/} $\nh$
in
\st. Since it is our goal to analyse an initial value problem in which a null
hypersurface $\nh$ is given, we shall {\em not\/} assume that
(\ref{(B5.7)}) holds unless explicitly specified otherwise.\footnote{{\em
Cf.\ }also \cite{HT,H} for an analysis in which (\ref{(B5.7)}) is not
assumed to hold, motivated by rather different considerations. The function
$e^H$ here corresponds precisely to the function $p_0 = \lim_{r\rai}\; p$
of \cite{HT}.} (We will, however, find it useful to impose  (\ref{(B5.7)})
when discussing the physical properties of the four dimensional
\st. This is clearly justified by the results of Section \ref{coordinates}.)

Let us show that polyhomogeneity of $h_{ab}$ implies that of $\beta, U^a,
V$ and $\frac{\pt h_{ab}}{\pt u}$:

\begin{Proposition}:
\label{P1}
Given any sequence $\{N_i\}_{i=0}^\infty$, $N_0 = 0$, there exists a
sequence $\{{\tilde N}_i\}_{i=0}^\infty$ with ${\tilde N}_0 = 0$, ${\tilde
N}_1 = N_1$, ${\tilde N}_ i \ge N_i$, such that for all $h_{ab} \in {\cal
A}^{\{{\tilde N}_i\}} \cap C^0(\bnh)$ satisfying $\lim_{r\rai}\; h_{ab}\,
dx^a\, dx^b = d\theta^2 + \sin^2\theta\, d\varphi^2$ we have
\begin{enumerate}
\item
\begin{equation}
\beta, U^a, r^{-2} V \in {\cal A}_{phg} \cap C^0(\bnh)\,;
\label{((B5.8))}
\end{equation}
\item If moreover
\begin{equation}
\lim_{r\rai}\; { U}^a = 0
\label{(B5.8.1)}
\end{equation}
holds, then we have, for any $j\ge 0$,
\begin{equation}
r^{-1} V \in {\cal A}_{phg} \cap C^0(\bnh)\,,
\label{(B5.8.2)}
\end{equation}
\begin{equation}
\left(\frac{\pt}{\pt u}\right)^j h_{ab} \in {\cal A}^{\{{\tilde N}_i\}}
  \cap C^0(\bar \nh)\,.
\label{(B5.8.3)}
\end{equation}
\end{enumerate}
\end{Proposition}

\noindent{\bf Remarks}.
1.
It is rather clear from the gauge character of
$\lim_{r\rai}\, { U}^a$ that (\ref{(B5.8.1)}) is not necessary for
(\ref{(B5.8.3)}) to hold, with possibly a different sequence $\{{\tilde
N}_i\}$. We have indeed verified this explicitly in the axisymmetric case
(the validity of (\ref{(B5.8.3)}) is in this case guaranteed by a cancellation
of some ``dangerous terms" which occur in the equation for $\frac{\pt
h_{ab}}{\pt u}$).  It should be noted that, assuming the initial
value problem for polyhomogeneous data on $\bar \nh$ can be solved, the
$u$-dependent terms arising when $\lim_{r\rai}\, { U}^a \neq 0$ would
lead to a $u$-dependent limit of $h_{ab}$
differing from the round metric on surfaces other than $\bar \nh$. This
can also be seen from inspection of the transformation used  above to remove
$X^a$. It does not affect the Proposition above which applies
only on $\bar \nh$. \\
2. Assuming the
characteristic initial value
problem for polyhomogeneous data can be
solved in the space of space--times with a polyhomogeneous Scri,
one way of understanding the significance of the relation
between the  $\{N_i\}_{i=0}^\infty$ and
 $\{{\tilde N}_i\}_{i=0}^\infty$
sequences is that
if we are given initial data in a
space characterized by the sequence $\{N_i\}_{i=0}^\infty$, then the
sequence $\{{\tilde N}_i\}_{i=0}^\infty$ can be chosen to be the one
appropriate to the evolution of that initial data. Sequences
 $\{{\tilde N}_i\}_{i=0}^\infty$
characterize those spaces which are invariant under evolution governed
by the vacuum Einstein equations.

\noindent{\bf Proof}: Replacing $h_{ab}$ by $\frac{\sin \theta}{\sqrt{\det
h_{ab}}}\, h_{ab}$ and $r$ by $r \frac{\sqrt{\det h_{ab}}}{\sin \theta}$
we may without loss of generality assume $\sqrt{\det h_{ab}} = \sin
\theta$.
A simple but somewhat tedious analysis of the
equations of Appendix \ref{vdbeqs}, making use of Proposition
\ref{PA1} in Appendix \ref{conventions}, gives the following:
the limits
$
\lim_{r\rai}\; \beta,\quad \lim_{r\rai}
{ U}^a,
$
exist and are, respectively, a smooth function and a smooth vector
field on a
sphere. We define
$$ H \equiv  \lim_{r\rai}\; \beta \ , \qquad
 X^a \equiv  -\lim_{r\rai}
{ U}^a\ ,
$$
\be
\label{(B5.11)}
\psi^a \equiv 2 e^{2H} {\cal D}^a H\, ,
\ee
where ${\cal D}^a$ is the covariant derivative with respect to
 the metric ${\hat h} =
\lim_{r\rai}\; h_{ab}\, dx^a\, dx^b$ on $S^2$.
We then have
\be
\beta - H  \in \frac{{\cal
A}_{phg}}{r^2}\,,\label{(B5.9)}
\ee
\be
{ U}^a + X^a - \frac{\psi^a}{r} \in \frac{{\cal
A}_{phg}}{r^2}\,,\label{(B5.10)}
\ee
\be
r^{-2}\, V \in {\cal A}_{phg} \cap C^0(\bnh)\,.
\label{(B5.12)}
\ee
Let us also define
\be
\psi \equiv e^{2H} \left(1 + 2\Delta_{\hat h} H + 4 |{\cal D} H|_{\hat
h}^2\right)\,,\label{(B5.14)}
\ee
where $\Delta_{\hat h}$ is the Laplacian of the metric ${\hat h}$, and
$|\cdot|_{\hat h}$ denotes the norm in ${\hat h}$.
If moreover $\lim_{r\rai}\;
{ U}^a = 0$, it follows that
\be
V - \psi r \in {\cal A}_{phg}\,,\label{(B5.13)}
\ee
\be
\frac{\pt h_{ab}}{\pt u} \in \frac{C^\infty(S^2)}{r} + \frac{{\cal
A}_{phg}}{r^2}\,,\label{(B5.15)}
\ee
In the above analysis the only not entirely
trivial step is to prove (\ref{(B5.15)}). Indeed, equations
(\ref{vdb5}) and (\ref{vdb6})
take the form
\begin{eqnarray}
\frac{\pt \phi_1}{\pt r} + f \phi_2 &= &\zeta_1\,,\label{(B5.16)}\\
\frac{\pt \phi_2}{\pt r} - f \phi_1 &= &\zeta_2\,,\label{(B5.17)}
\end{eqnarray}
where, using the notation of \cite{vdB} and Appendix \ref{vdbeqs},
\begin{eqnarray*}
\phi_1 &= &r \cosh(2\delta)\, \frac{\pt\spm}{\pt u}\,,\\
\phi_2 &= &r\, \frac{\pt\delta}{\pt u}\,,\\
f &= &2 \sinh( 2\delta)\, \frac{\pt\gamma}{\pt r}\,,
\end{eqnarray*}
and the $\zeta_a$, $a = 1, 2$, can be found in Appendix \ref{vdbeqs}. The
hypothesis $h_{ab} \in {\cal A}^{phg} \cap C^0(\bnh)$
is equivalent to $\spm, \delta \in {\cal A}^{phg} \cap
C^0(\bnh)$, so that from (\ref{(B5.9)})--(\ref{(B5.14)}) one finds
\be
\label{cond:new}
rf,
\  \zeta_a \in \frac{{\cal A}_{phg}}{r^2}\,, \qquad a = 1, 2\,.
\end{equation}
It is an exercise in ODE's to show
that under (\ref{cond:new}) any solution of (\ref{(B5.16)})--(\ref{(B5.17)})
is necessarily polyhomogeneous, with
$$
\phi_a \in C^\infty(S^2) + r^{-1}\,
{\cal A}_{phg}\ .
$$
[The result can be proved by {\em e.g.\ }setting up a contraction principle
argument in weighted spaces of the kind used in Appendix
\ref{geodesics}.]
This implies $\frac{\pt h_{ab}}{\pt u} \in r^{-1}\,
C^\infty(S^2) + r^{-2}\, {\cal A}_{phg}$, as claimed.

To prove our claim about existence of {\em self--consistent\/} sequences
$\{{\tilde N}_i\}$ which have the property that $h_{ab} \in
{\cal A}^{\{{\tilde N}_i\}}$ is {\em formally} preserved under time
evolution, define
$$N = N_1\,,$$
so that
$$|h_{ab} - \hat h_{ab}| = O(r^{-1}\, \log^N r)\,.$$
Consider a term, say $\chi$, in $h_{ab}$ which has a radial behaviour
$r^{-i}\, \log^j r$. From the equations of Appendix \ref{vdbeqs}
one easily finds that
such a term produces terms $r^{-i-1}\, \log^{N+j} r + \mbox{lower
order}$\footnote{By ``lower order" we mean terms which have the same power
of $r^{-1}$ and smaller
powers of $\log r$, or higher powers of $r^{-1}$.} in $\beta$. Next, if $i =
2$, such a term will produce terms
$r^{-3}\, \log^{j+1} r +
\mbox{lower order}$ in ${ U}^a$, while for $i \ne 2$ it will lead to
terms
$r^{-i-1}\, \log^{j} r + \mbox{lower order}$ in ${ U}^a$.
Using the Kronecker $\delta_a^b$ defined as usual by
 $$
a,b\in\R~, \qquad \delta_a^b=\cases{1, & for $a=b$, \cr 0,& otherwise, }
 $$
we conclude that $\chi$ produces terms
 $r^{-i-1}
\log^{j+\delta_i^2} r + \mbox{lower order}$ in ${ U}^a$.
There is a cancellation in the equation for $V$ which implies that
if $i=1$ and $N(=j)=0$,
$\chi$ will generate terms
$r^{-1}\, \log^{{\hat N}_1} r + \mbox{lower order}$ in $V$, rather than
the $\log r + \mbox{lower order}$ terms which would have appeared if the
cancellation had not taken place: here ${\hat N}_1$ is determined by
$N_2$ (${\hat N}_1 = 0$ if $N_2 = 0$). In general, the cancellation implies
that the leading term in $V$ arising from
$i = 1$ and $N > 0$ is of order $\log^N r$, while  the leading
contribution\footnote{The radial behaviour $r^{1-i}\log^{j+\delta_i^2}
r$ is obtained here by a {\sc sheep} calculation. It seems that
knowledge of some cancellations for $i=1$ is necessary for the
argument to hold. Nevertheless for $i\ge 1$
a straightforward analysis
of the $V$ equation
(\ref{vdb4}) yields  a leading order
contribution $r^{1-i}\log^{j+\delta_i^1+\delta_i^2}r $ to $V$ from
$\chi$, without going into
the details of the cancellation structure of the equations (which {\sc
sheep} automatically does).
One could use this estimate of the contributions of $\chi$ to $V$ in
the remainder of the argument to prove Proposition \ref{P1}, with a
perhaps somewhat ``worse'' sequence ${\tilde N}_i$.}
 from $\chi$ is of order $r^{1-i}\log^{j+\delta_i^2}
r$.
Inserting all this information in the equations for $\frac{\pt
h_{ab}}{\pt u}$ one in general obtains a contribution $r^{-1-i}\,
\log^{j+\delta_i^2} r + \mbox{lower order}$
from $\chi$, but an $i=1$, $ N > 0$
term contributes\footnote{This  exponent
is, again, obtained using {\sc sheep}. Analytically it is more or less
straightforward  to
estimate the right hand side of (\ref{(B5.18)}) as $O(r^{-2}\,
\log^{N_1+1} r)$. Such an estimate would be sufficient for the main
conclusion to remain valid.}
  only a term of order
$r^{-2}\log^{N-1} r$.

To construct
a sequence $\{{\tilde N}_i\}$ given a sequence $\{N_i\}$,
set ${\tilde N}_1 = N_1$. Then from the equations for $\frac{\pt
h_{ab}}{\pt u}$ one obtains, if $N_1 > 0$,
\begin{equation}
\frac{\pt}{\pt u} \left(h_{ab} - \frac{h_{ab}^1(u, x^a)}{r}\right) =
O\left(r^{-2}\, \log^{N_1-1} r\right) \ ,
\label{(B5.18)}
\end{equation}
for some
$h_{ab}^1(u, x^a)$.
If $N_1 = 0$, the next contributions to the $u$--derivatives of the
$\gamma$ and $\delta$ used in Appendix \ref{vdbeqs} are $O(r^{-3}
\log^{N_2+1} r)$,
which implies that the time derivative of the trace--free part of
$h_{ab}-h_{ab}^1(u, x^a)/r$ is $O(r^{-3}
 \log^{N_2+1} r)$;
 on the other hand, the time derivative of the ``trace part'' of
$h_{ab}-h_{ab}^1(u, x^a)/r$ is $O(r^{-2})$ and is determined uniquely by the
$u$--derivatives of
$h_{ab}^1(u, x^a)$.
(\ref{(B5.18)}) shows that the coefficients of the $r^{-1}\, \log^i r$, $i =
1, \ldots, N_1$ are constants of motion, and if we set ${\tilde N}_2 =
\max(N_2, N_1-1)$ then the space ${\cal A}^{\{{\tilde N}_i\}}$ will be
formally preserved by evolution up to $O(r^{-3+\epsilon})$, $\epsilon
>0$,
 terms. Proceeding
recursively one can construct a
self--consistent sequence $\{{\tilde N}_i\}$.
The analysis of the
higher
$u$--derivatives proceeds in a similar manner by considering the
equations
obtained from the Einstein equations by $u$--differentiation, and the result
follows.\hfill$\Box$

{}From what has been said in the proof above it should be clear that if
we write, along $\nh$,
$$
h_{ab}\sim
\sum_{i=0}^\infty\sum_{j=0}^{N_i}h_{ijab}(\theta,\varphi)\,r^{-i}\log ^j r\ ,
$$
then on $\nh$ from the Einstein equations one obtains a Bondi -- van der
Burg -- Metzner type hierarchy of
equations
$$
\frac{\partial h_{ijab}}{\partial u} = F_{ijab}\ ,
$$
where $F_{ijab}$ is a function of $\theta,\,\varphi$
and the $h_{k\ell cd},\, 0\le k \le i-1$ together with a finite number of
their derivatives. In
particular, if we assume that a polyhomogeneous expansion of the
metric also holds in a neighbourhood of $\nh$ one obtains:

\begin{Proposition}:
\label{P2}
Under the hypotheses of Proposition \ref{P1}, including (\ref{(B5.8.1)}),
we have:
\begin{enumerate}
\item The coefficients of $r^{-1}\, \log^j r$, $1 \le j \le N_1$, in $h_{ab}$
are constants of motion.
\item Suppose that $N_1 = 0$.
Then the coefficients of $r^{-2}\, \log^j r$, $1 \le j \le
{\tilde N}_2 = N_2$ in $h_{ab}$
and
 the coefficients of  $r^{-2}$ in
the trace--free part of the
 $h_{ab}$ are constants of motion.
\item Suppose that on $\nh$ we have
\begin{equation}
h_{ab} - \lim_{r\rai}\; h_{ab} = \frac{h_{ab}^1(x^a)}{r} +
\frac{a(x^a)\breve h_{ab}(x^a)}
{r^2}+
O\left(r^{-3}\, \log^{N_3} r\right)\,,\label{(B5.19)}
\end{equation}
with $\breve h_{ab}$ given by \eq{roundmetric},
and that $N_1 = \ldots =N_i = 0$, $i \ge 2$. Then we can set $N_j = {\tilde
N}_j$, $j = 0, \ldots, i+1$ and the coefficients of $r^{-i-1}\, \log^j
r$, $1 \le j \le N_{i+1} = {\tilde N}_{i+1}$ in $h_{ab}$ are constants
of motion.
\end{enumerate}
\end{Proposition}

{\bf Remarks:} In the notation of \cite{vdB}, \eq{(B5.19)} is
equivalent to
$\gamma = c/r+O(r^{-3}\log^{N_3}r)$, $\delta = d/r+O(r^{-3}\log^{N_3}r)$.
Note that the time dependence of the $r^{-1}$ contributions to
$h_{ab}$ is undetermined by the above considerations, which are
purely asymptotic\footnote{These
contributions define
an analogue of what is called a   ``news function'' in
\cite{BMS,Sachs}, and should be determined by the behaviour of
sources and/or of the gravitational field in the interior,
or perhaps by some interior boundary conditions,
in a complete solution.}. In Section 4 we show that the only remaining
coordinate freedom is given by the BMS group, which implies that we
have only one arbitrary function of $\theta$ and $\phi$ available for
changing the values of the constants of motion just obtained, so that
not more than one of them (if any) can be set to a fixed value.

It is natural to consider extending the second result in the Proposition
and ask what is the ``smallest'' self--consistent
sequence ${\tilde  N}_i$ if $N_i = 0$ for all $i$
and it is not assumed {\it a priori} that
(\ref{(B5.19)}) holds. When
$\lim_{r\rai}\; U^a = 0$, one easily obtains, from what
has been said earlier, that we can set
$${\tilde N}_0 = {\tilde N}_1 = {\tilde N}_2 = 0$$
and
$$
{\tilde N}_3= 1\ .
$$
Moreover in such a case the arguments of the proof of Proposition \ref{P1}
show that we will have
\cmm{I think the first two of these are both wrong and in disagreement
with the results above: the smallest ones should be $N^U_3=1$ and
$N^V_1=1$.}
\begin{eqnarray*}
\beta &\in &{\cal A}^{\{N_i^\beta\}}\,, \qquad N_0^\beta = N_1^\beta =
N_2^\beta = N_3^\beta  = 0\,, \qquad N_4^\beta = 1\,, \\
{ U}^a &\in &{\cal A}^{\{N_i^U\}}\,, \qquad N_0^U = N_1^U =
N_2^U = 0\,, \qquad N_3^U = 1\,,\\
V - r &\in &{\cal A}^{\{N_i^V\}}\,, \qquad N_0^V =0\,, \qquad N_1^V = 1\,.
\end{eqnarray*}
In asymptotically Minkowskian coordinates $$(t, x, y, z) = (u+r,\, r
\sin \theta \cos \phi,\, r \sin \theta \sin \phi,\, r \cos \theta)$$
this corresponds to a metric
(\ref{(B5.1)}) which along $\nh$ approaches the Minkowski one  as
$$\spm_{\mu\nu} - \eta_{\mu\nu} = \frac{\spm_{\mu\nu}^1}{r} +
\frac{\spm^{2,1}_{\mu\nu}\, \log r}{r^2} + \frac{\spm_{\mu\nu}^2}{r^2} +
O(r^{-2-\epsilon})\,, \quad \epsilon > 0\,.$$

The results of Proposition \ref{P1} can be generalized to include both
matter fields and rather weaker asymptotic conditions. Let us start
by defining a space of functions
$\C{\mu,\lambda}$: for $\mu,\lambda\in\R$ a function $f$ will be said to be
in
$\C{\mu,\lambda}(\nh)$, or for short in $\C{\mu,\lambda}$,
 if for all $i\in \N $ and for all multi--indices
$\alpha$ we have
$$
|\partial_r^i\partial_v^\alpha f |\le  C_{i,\alpha}r^{-\mu-i}(1+|\log
r|)^{\lambda}\ ,
$$
for some constants $C_{i,\alpha} $, where the coordinates $v$ stand
for
$\theta,\phi$. In order to  avoid in what follows a rather
annoying  discussion of some not so interesting special cases we shall
always assume that in all spaces considered
if $\mu=1,2,3$ or $4$ then the logarithmic behaviour
exponent $\lambda$ satisfies $\lambda\ne -2,-1$.
 We wish to show that the space of metrics on $\nh$ of
the form
\be
\label{NCOND}
h_{ab}-\breve h_{ab} \in
r^{-1}\A+\C{\mu,\lambda}\ , \qquad \mu > 0
\ee
is {\em formally} preserved by evolution with the Einstein equations with
possibly some matter fields.
Let us
 denote   by
$\lhs_n$, respectively by $\rhs_n$, the left hand side, respectively
the right
hand side, of the $n$'th
equation of Appendix \ref{vdbeqs}. Then, in the presence of matter the
Einstein equations become
\be
\label{meq.1}
\lhs_1=\rhs_1 + \frac{r}{4}\hat T_{11}\ ,
\ee
\be
\label{meq.2}
\lhs_2=\rhs_2 + 2r^2 \hat T_{12}\cosec \theta  \ ,
\ee
\be
\label{meq.3}
\lhs_3=\rhs_3 + 2r^2 \hat T_{13} \ ,
\ee
\be
\label{meq.4}
\frac{\partial V}{\partial r} = \lhs_4=\rhs_4 -
\frac{e^{2\beta}h^{ab}}{2}\hat T_{ab}\ ,
\ee
\be
\label{meq.5}
\lhs_5=\rhs_5 +\frac{e^{2\beta}}{4r}
\left(e^{-2\gamma}\hat T_{22} -e^{2\gamma}\sin^{-2}\theta\,\hat T_{33}
\right)\ ,
\ee
\be
\label{meq.6}
\lhs_6=\rhs_6 +\frac{e^{2\beta}}{2r\cosh 2\delta \sin \theta }
\left(\hat T_{23}
-\frac{1}{2}h^{cd}\hat T_{cd}h_{23}\right)\ ,
\ee
where $h^{ab} $ is the matrix inverse to $h_{ab}$,
$$
\hat T_{\mu\nu} = \kappa (T_{\mu\nu}-\frac{T}{2}\gamma_{\mu\nu})\ ,
$$
$\kappa$ is the gravitational coupling constant, and $T_{\mu\nu}$ is
the energy momentum tensor of the matter fields. We shall require
\begin{eqnarray}
\label{mcond.1}
\hat T_{11 } \in r^{-3 } \A + \C{\mu_{11 },\lambda_{11 }},
& \mu_{11 } >2 \ ,\\
\label{mcond.2}
\hat T_{1a } \in r^{-2 } \A + \C{\mu_{1a },\lambda_{1a }},
\quad a=2,3,
&
\mu_{12 }=\mu_{13} >1 \ , \quad\lambda_{12}=\lambda_{13}\ ,
\\
\label{mcond.3}
h^{ab}\hat T_{ab } \in r^{-1 } \A + \C{\mu_{0 },\lambda_{0 }},
&
\mu_{0 } >0 \ ,
\\
\label{mcond.4}
e^{2\gamma}\sin^{-2}\theta\,\hat T_{33}
-e^{-2\gamma}\hat T_{22}
\in r^{-1 } \A + \C{\mu_{r },\lambda_{r }},
& \mu_{r } > 0\ ,
\\
\label{mcond.5}
\hat T_{23}
-\frac{1}{2}h^{cd}\hat T_{cd}h_{23}
\in r^{-1 } \A + \C{\mu_{r },\lambda_{r }},
&
\end{eqnarray}
where the various powers in front of $\A$ and the exponents in
$\C{\mu_*,\lambda_*}$ have been chosen so that the leading order behaviour
of the various functions which appear in the metric coincides with
the behaviour one observes in the vacuum case. Assuming that
$\lim_{r\rightarrow\infty}\beta=\lim_{r\rightarrow\infty}U^a=0$,
an analysis as described in
the proof of Proposition \ref{P1} leads to
$$
\frac{\partial}{\partial u}h_{ab} \in r^{-1} \A +
\C{\mu+1,\lambda+\delta_{\mu}^2}+
\C{\mu_{11}-1,\lambda_{11}+\delta_{\mu_{11}}^4} +
\qquad\qquad\qquad\qquad\qquad
$$
\be
\label{TCOND}
\qquad\qquad+
\C{\mu_{1a},\lambda_{1a}+\delta_{\mu_{1a}}^3} +
\C{\mu_r+1,\lambda_r} + \C{\mu_0+2,\lambda_0+\delta_{\mu_{0}}^1+N_1}\ .
\ee
It follows that for
$$
\mu\le\min\{\mu_{11}-1,\mu_{1a},\mu_r+1,\mu_0+2\}
$$
(with appropriate inequalities for the $\lambda_*{}$'s if the above
inequality is an equality) the Einstein equations will {\em formally}
preserve the
space of metrics $h_{ab}$ satisfying (\ref{NCOND}).
The proof of (\ref{TCOND}) follows immediately by integration of the
equations
(\ref{meq.1})--(\ref{meq.6}), and of the equations which are obtained
by $u$--differentiation of those. Step by step one obtains:
$$
\beta \in r^{-1}\A+ \C{\mu+1,\lambda+N_1} +
\C{\mu_{11}-2,\lambda_{11}}\ ,
$$
$$
U^a \in r^{-2}\A+\C{\mu+1,\lambda+\delta_\mu^2}
+\C{\mu_{11}-1,\lambda_{11}+\delta_{\mu_{11}}^4}
+\C{\mu_{1a},\lambda_{1a}+\delta_{\mu_{1a}}^3}
\ ,
$$
$$
V-r \in \A +
\C{\mu-1,\lambda+\delta_\mu^1+\delta_\mu^2}+
\C{\mu_0-1,\lambda_0+\delta_{\mu_{0}}^1}
+\C{\mu_{11}-3,\lambda_{11}+\delta_{\mu_{11}}^3+\delta_{\mu_{11}}^4}
+\C{\mu_{1a}-2,\lambda_{1a}+\delta_{\mu_{1a}}^2+\delta_{\mu_{1a}}^3}
\ ;
$$
the above equations inserted in the evolution equations
(\ref{meq.5})--(\ref{meq.6}) yield (\ref{TCOND}).
[If we assume that the
appropriate decay properties
are also satisfied by $\partial^j\hat T_{\mu\nu}/\partial u^j$ on $\nh$,
$j=0,\ldots,J$,
then a preliminary examination suggests that
$\partial^i h_{ab}/\partial u^i$ will also be of the form
(\ref{TCOND}) for $i=0,\ldots,J$.]
  We note that the
previous calculations on similar lines
\cite{CouchTorrence,NovakGoldberg} led to upper bounds on the
parameter $\mu$ which do not appear here because, unlike those previous
calculations, we do not forbid the appearance of $\log r$ terms.

\section{Geometric Interpretation}
\label{interpretation}

Consider a metric of the form (\ref{(B5.1)}). Following Penrose
\cite{Penrose} it is useful to introduce a new coordinate
$$x \equiv r^{-1}\,,$$
so that by Proposition \ref{P1} when $h_{ab} \in {\cal A}^{phg} \cap
C^0(\bnh)$ the metric
\begin{eqnarray}
{\tilde\gamma}_{\mu\nu}\, dx^\mu\, dx^\nu &\equiv &x^2\,
\gamma_{\mu\nu}\, dx^\mu\, dx^\nu\nn\\
&= & -V x^3\, e^{2\beta}\, du^2 + 2e^{2\beta}\, du\, dx
+ h_{ab}(dx^a + U^a\, du) (dx^b + U^b\, du)\label{(B5.20)}
\end{eqnarray}
is {\em polyhomogeneous\/} on the set $\{x \in [0, 1/R],\, x^a \in S^2\}$,
{\em i.e.\/}, there exists a sequence $\{{\hat N}_i\}$ and
functions ${\tilde
\gamma}_{\mu\nu ij}(x^a) \in C^\infty (S^2)$ such that
$$
{\tilde \gamma}_{\mu\nu} \sim \sum_{i=0}^\infty\, \sum_{j=0}^{{\hat
N}_i}\, {\tilde \gamma}_{\mu\nu ij}(x^a)\, x^i\, \log^j\, x\,,$$
$${\hat N}_0 = 0\,.
$$
(When (\ref{(B5.19)}) holds and $h_{ab} \in C^\infty(\bnh)$ one actually
obtains ${\hat N}_i = 0$ for all $i$, and the metric (\ref{(B5.20)}) is in
$C^\infty(\bnh)$. This corresponds to the standard Bondi--Penrose--Sachs
situation of a smooth \Scri.)\ We have the following result, which is
established by calculating the Christoffel symbols of the metric
$x^2\gamma_{\mu\nu}$ using {\sc sheep}, and making use of vacuum Einstein
equations  for $\gamma_{\mu\nu}$ in the first leading orders:

\begin{Proposition}:
%
%
\label{P3}
Consider a characteristic initial data set with $h_{ab}\in {\cal
A}_{phg}\cap C^0({\bnh})$,
$\lim_{r\rightarrow \infty}h_{ab}=d\theta^2+\sin^2\theta\,d\varphi^2$
and\/
 $\lim_{r\rightarrow \infty}U^a=0$,
and set $\Omega = x = r^{-1}$.
We have
\begin{equation}
\lim_{x\raz}\; \tilde\nabla_a  \tilde\nabla_b \Omega = 0\,, \qquad a,
b = 2, 3\,,
\label{(B51.1)}
\end{equation}
where $\tilde\nabla$ is the covariant derivative of the metric $\tilde
\gamma_{\mu\nu} \equiv x^2\gamma_{\mu\nu}$.
If moreover $\lim_{x\raz}\; \beta = 0$, then we also have
\begin{equation}
\lim_{x\raz}\; \tilde\nabla_\mu \tilde
\nabla_\nu \Omega = 0\,, \qquad \mu, \nu = 0,
\ldots, 3\,.
\label{(B51.2)}
\end{equation}
\end{Proposition}
Recall that the geometric meaning of (\ref{(B51.1)}) is the vanishing of the
shear of the hyper\-surface $\{x = 0\}$ ({\em cf.\ e.g.\ }\cite{Wald}).
Another interpretation of (\ref{(B51.1)}) is that the {\em conformal extrinsic
curvature\/} of \Scri\ vanishes, {\em cf.\ }\cite{ACh}[Appendix]. This
property of \Scri\ is well known for ${\tilde \gamma}_{\mu\nu} \in
C^\infty(\bnh)$. It should be stressed that for general polyhomogeneous
metrics as constructed (on $\nh$) in the previous section, the fact that the
left hand sides of (\ref{(B51.1)})--(\ref{(B51.2)}) exist and are bounded is
a non--trivial statement which {\em makes use\/} of vacuum Einstein
equations to the first two leading orders,
because equations
(\ref{(B51.1)})--(\ref{(B51.2)}) contain derivatives of the metric which
could potentially blow up as $\log^{{\hat N}_1} x$ as $x\raz$. Proposition
\ref{P3} is the original observation which led to the proof in
\cite{ACWeyl}, that generic Cauchy data constructed by the conformal
method as in \cite{ACh} will lead to \sts\ which {\em cannot\/} admit a
polyhomogeneous \Scri.

Throughout the remainder of this section we shall assume that
\be
\label{matdef}
\lim_{r\rightarrow \infty} h_{ab}dx^adx^b=d\theta^2 +\sin^2\theta\,
d\varphi^2 =:\breve h_{ab}dx^adx^b\ ,
\ee
\be
\label{matdef1}
\lim_{r\rightarrow \infty}\beta=0, \qquad
\lim_{r\rightarrow \infty}U^a=0\ .
\ee

A textbook property of smooth \Scri's is that the Weyl tensor of
the conformally rescaled metric vanishes at \Scri, {\em cf.\ e.g.\
}\cite{Wald}.
A {\sc sheep} calculation of the Weyl tensor of the metric ${\tilde
\gamma}_{\mu\nu}\, dx^\mu\, dx^\nu$,
assuming that $\gamma_{\mu\nu}$ is vacuum,
gives:

\begin{Proposition}:
%
%
\label{P4}
In addition to  the hypotheses of Proposition \ref{P3}, let
$h_{ab} \in C^2(\bnh)$, and let (\ref{matdef})--(\ref{matdef1}) hold.
In local coordinates define
$$
\chi_{ab} = \lim_{x\raz}\;\left( \frac{\pt^2 h_{ab}}{\pt
x^2} - \frac{1}{2} h^{cd}\frac{\pt^2 h_{cd}}{\pt
x^2}\,h_{ab}\right)\ .
$$
 Let $\tilde C_{\alpha\beta\gamma\delta}$ denote the components of the
Weyl tensor of $\tilde \gamma$ in the half--null tetrad
$$
\theta^0=e^{2\beta}du
\ ,\qquad
\theta^1=\frac{x^3V}{2}du - dx\ ,
$$
$$
\theta^2=-(\cosh \delta \,e^\gamma\,U^\theta +\sin \theta\, \sinh \delta\,
e^{-\gamma} U^\phi)\,du+\cosh \delta \,e^\gamma \,d\theta
+\sin\theta\,\sinh\delta\, e^{-\gamma}\, d\varphi \ ,
$$
\be
\label{tetrad}
\theta^3=-(\sinh \delta \,e^\gamma\,U^\theta +\sin \theta\, \cosh \delta\,
e^{-\gamma} U^\phi)\,du +\sinh \delta \, e^\gamma d\theta +
\sin\theta\,\cosh \delta\, e^{-\gamma}\,d\varphi\ .
\ee
Then we have
\begin{equation}
\lim_{x\raz}\; {\tilde C}_{1a1b} = \chi_{ab}\ ,
\label{(6I.1)}
\end{equation}
while all the remaining components of $\lim_{x\raz}\; {\tilde
C}_{\alpha\beta\gamma\delta}$ vanish, except of course those which
can be obtained by appropriate permutations of ${\tilde C}_{1a1b}$.
\end{Proposition}

[Let us point out that Proposition \ref{P4} will still be true with
non-vanishing Ricci curvature provided that (\ref{NCOND}) and
(\ref{mcond.1})--(\ref{mcond.5}) hold with $\mu_0,\mu_r>1$,
$\mu_{1a}>2$, $\mu, \mu_{11}>3$, and that the  powers of $r^{-1}$
in front
of the polyhomogeneous pieces in $\hat T_{\mu\nu}$ in eqs.\
(\ref{mcond.1})--(\ref{mcond.5}) are increased by one.]
Proposition \ref{P4} shows that the Weyl tensor of ${\tilde
\gamma}_{\mu\nu}$ vanishes at $x = 0$ if and only if the trace free
part of $\frac{\pt^2
h_{ab}}{\pt x^2}\big|_{x=0}$ vanishes, {\em i.e.}, if and only if what
Bondi {\em et al.\ }termed the
``outgoing radiation condition", (\ref{(B5.19)}), holds.
Let us also
mention that (\ref{(6I.1)}) is equivalent to
$$
\tilde \Psi_0\Big|_{x=0} = \tilde \Psi_1\Big|_{x=0} = \tilde
\Psi_2\Big|_{x=0} = \tilde \Psi_3\Big|_{x=0}  = 0, \qquad
\tilde \Psi_4\Big|_{x=0}\ne0\ ,
$$
where the $\tilde \Psi_i$'s are the Newman--Penrose components of the
Weyl tensor of the metric $\tilde\gamma$
in a null tetrad related to the tetrad (\ref{tetrad})
above in the obvious way.

It is worthwhile emphasizing that it follows from Proposition \ref{P2}, point
2, and from (\ref{(6I.1)}) that the components ${\tilde
C}_{1a1b}(u\,,0\,,x^a)$ are {\em pointwise constants of motion\/},
{\em i.e.},
independent of $u$. This result seems to have been
already observed by Winicour \cite{Winicour}, and independently
by Christodoulou and Klainerman \cite{ChKl} (under
much weaker asymptotic conditions).

Perhaps the most important result of the Bondi--Sachs analysis is the well
known theorem
(originally due to Trautman \cite{Trautman})
that Bondi's mass  is a
decreasing function of $u$.
For the metrics under consideration here, let us define the Bondi mass as
($-1/2$) the integral over the
sphere $S^2$ of the $r^0$ coefficient in the expansion of $V$. This
definition clearly reduces to the original one by Bondi--Sachs, when
the conditions imposed by Bondi and Sachs hold.
We
have found that, under (\ref{matdef})--(\ref{matdef1}), and whatever the
sequence $\{N_i\}$ and the $h_{ab} \in {\cal A}^{\{N_i\}}$, the mass
loss equation (35) of
\cite{BMS}, which can be obtained by equating to zero the integral
over $S^2$ of the right side of (\ref{bondimassloss}),
({\em cf.\ }also \cite{Sachs}, eq.\ (4.16)) remains
unchanged. This can be seen as follows: under the
above conditions it follows from the vacuum Einstein equations that
$$
V-r,\  r^2\beta, \ r^2 U^a \in \A\ .
$$
As has been shown by Sachs \cite{Sachs} following the original
observation of Bondi {\em et al.\ }\cite{BMS}, in the coordinate
system
of (\ref{(B5.1)}) we have  (if the other field equations hold)
$$
R_{00}=0 \quad \Leftrightarrow \quad
\lim_{r\rightarrow\infty}r^2R_{00}=0 \ .
$$
A {\sc sheep} calculation gives
\be
\label{bondimassloss}
\lim_{r\rightarrow\infty}r^2R_{00}=
\lim_{r\rightarrow\infty}\left(\frac{\partial V}{\partial u} -
\frac{1}{4}r^4
h^{ab}h^{cd}\frac{\partial^2 h_{ac}}{\partial u \partial r}
\frac{\partial^2 h_{bd}}{\partial u \partial r}
-r^2{\cal D}_a\frac{\partial U^a}{\partial u}\right)\ ,
\ee
where ${\cal D}_a$ is the covariant derivative of the metric $\breve
h_{ab}$. It is clear from (\ref{bondimassloss}) and from what has been
said
before that those $\log$ terms which could contribute to this equation,
if any, drop out because their $u$ derivatives vanish.
It follows
that for {\em all polyhomogeneous\/}
initial data
the Bondi mass is a non--increasing function of time when\footnote{When
$\lim_{r\rai}\; \beta \ne 0$, then the ``mass aspect" is
no longer given by the standard Bondi--Sachs expression. But
monotonicity of an
appropriately defined total Bondi mass of course remains valid.}
 $\lim_{r\rai}\; \beta =
0$.

We would like to point out that it is clear that the mass as
defined
above is the
``correct mass'' for polyhomogeneous initial data
$h_{ab} \in {\cal A}^{\{N_i\}}$ with $N_0=N_1=0$. On the other hand
we believe that some care should be taken when interpreting the above
as the mass in the case $N_1\ne 0$, since in that case the leading
order
behaviour of $V-r$ is logarithmic, which might reflect an infinite or
ill defined
mass of the system.
Such a possibility is also suggested by
eq.\ (\ref{meq.4}) which shows that
 some $\log$ terms might arise
from $1/r$ terms in $h^{ab}\hat T_{ab}$. Now in an orthonormal tetrad
in which $e^0$ is {\em timelike} one finds that
  $h^{ab}\hat
T_{ab}=-\kappa \, r^2(T^0_0+T^1_1)=\kappa \, r^2(T_{00}-T_{11})$,
so that 1/r terms in $h^{ab}\hat
T_{ab}$ correspond to an {infinite} amount of matter energy: if
$r^2T_{00}$ behaves as $1/r$, then
 $T_{00}$ (the matter energy density)
ceases to be integrable
over $\nh$. A thorough analysis of the conditions under which
the mass at null infinity is finite and well defined lies outside the
scope of this paper.

It is natural to ask what happens with the ``peeling--off" property for \sts\
for which (\ref{(B5.3)}) fails to hold. A {\sc sheep} calculation shows that
(in the vacuum case)
for any $h_{ab} \in C^0(\bnh)\cap\A$  along
$\nh$ we
have
$$R_{\alpha\beta\gamma\delta} =
\frac{R_{\alpha\beta\gamma\delta}^1}{r} +
\frac{R_{\alpha\beta\gamma\delta}^2}{r^2} +
\frac{R_{\alpha\beta\gamma\delta}^{\log}\, \log r}{r^3} +
\frac{R_{\alpha\beta\gamma\delta}^3}{r^3} + \cdots\,,$$
with $R_{\alpha\beta\gamma\delta}^1$ and
$R_{\alpha\beta\gamma\delta}^2$ peeling off exactly in the same way as
they would if (\ref{(B5.3)}) were satisfied, {\em i.e.\
}$R_{\alpha\beta\gamma\delta}^1$ is of type N,
and $R_{\alpha\beta\gamma\delta}^{2}$ is of
type III; in fact
$R_{\alpha\beta\gamma\delta}^1$ and $R_{\alpha\beta\gamma\delta}^2$
are exactly the same as in the case considered by
Sachs \cite{Sachs}
({\em cf.\ }also \cite{Trautman}).
This follows from the fact that {\em all} terms which
contribute to $R_{\alpha\beta\gamma\delta}^1$ and
$R_{\alpha\beta\gamma\delta}^2$ are $u$--differentiated, and those $\log$
terms which could potentially contribute at this order are constants
of motion. Here $R_{\alpha\beta\gamma\delta}^{\log}$ may contain
powers of $\log r$.

In the case ${\tilde \gamma}_{\mu\nu} \in C^\infty(\bnh)$ it was observed
in \cite{NPCM,NPCM1} that
there exist some nontrivial global constants of motion for
a vacuum gravitating system. We wish to point out that these
quantities cease to be constants of motion even in the case in
which
$h_{ab}\in C^\infty(\nhb)$ if one does not assume
that (\ref{(B5.19)}) holds. Clearly it is sufficient to prove that
assertion for those
metrics (\ref{(B5.1)}) which are of the
Bondi--van der Burg--Metzner form \cite{BMS}:
\begin{eqnarray}
\frac{\pt\spm_{\mu\nu}}{\pt\varphi} &= &0\label{(C1)}\ ,
\\
h_{ab}\, dx^a\, dx^b &= &e^{2\spm}\, d\theta^2 + e^{-2\spm}\,
\sin^2\theta\, d\varphi^2\,,\label{(C2)}\\
U^\varphi &= &0\,.\label{(C3)}
\end{eqnarray}
Let us expand the function $\spm$
appearing in (\ref{(C2)})
as
$$\spm = \frac{c}{r} + \frac{\spm_2}{r^2} +
\frac{\spm_{3,1}\log r}{r^3} +
\frac{\spm_3}{r^3} +
\frac{\spm_{4,1}\log r}{r^4} +
\frac{D}{r^4} + \cdots\,.$$
[The terms $\spm_{3,1}$ and $\spm_{4,1}$ above are necessary:  even if
${\spm_{3,1}}\Big|_{u=0} = {\spm_{4,1}}\Big|_{u=0}=0$
 (which we are free to assume),
we shall have $\spm_{3,1}\ne 0 \neq \spm_{4,1}$ at later times in
general, as a consequence of the evolution equations.]
In the axisymmetric case with $\spm_2 =\spm_{3,1}\equiv 0$ the constant of
motion is given by \cite{vdB}
\begin{eqnarray*}
{\cal D} &= &\int_{S^2} D \sin^2 \theta\, d\mu_0\,,\\
d\mu_0 &= &\sin \theta\, d\theta\, d\phi\,.
\end{eqnarray*}
If one takes the minimal sequence of $\tilde{N}_i$ as described in
section 2 and allows $\spm_2 \not\equiv 0$, one finds that
the  $\spm_{3,1}$ and  $\spm_{4,1}$  terms
(but not $r^{-3}\log^j r$ or $r^{-4}\log^j r$  with $j>1$) arise. After a
long calculation it turns out that the $u$--derivative of the $\spm_{4,1}$
term, when multiplied by $\sin^3 \theta$,
is a total derivative with respect to $\theta$ and,
removing
further total derivative terms, one finds that the equation of
motion for ${\cal D}$ takes the form
$$\frac{\pt {\cal D}}{\pt u} = \int_{S^2} G d\theta\,d\phi $$
where
\begin{eqnarray}
G & =&
\bigg\{\gamma_2 \sin \theta \Big[\sin^2 \theta ( - 4
\frac{\partial^2c}{\partial \theta^2}
          + 16 M - 10 u + 4 c)
         - 24 \sin \theta  \cos \theta\frac{\partial c}{\partial \theta}
         - 16 \cos^2 \theta \, c\Big] \nn \\
&& \quad         + 15 \sin^3 \theta\, \gamma_{3,1}^0\bigg\}/12,
\label{timederivative stuff}
\end{eqnarray}
$\gamma_{3,1}^0 \equiv {\gamma_{3,1}}\Big|_{u=0}$,
  and $M$ is the usual Bondi mass aspect
({\em i.e.}, $-2M$ is the integration constant which appears when integrating
the $V$ equation (\ref{vdb4})).

For given $M\Big|_{u=0}\not\equiv 0 $, $\gamma_{3,1}\Big|_{u=0} $
(perhaps, but not necessarily, being zero) and
$c\Big|_{u=0}$
it seems obvious  from \eq{timederivative stuff}
that
the function $\spm_2\Big|_{u=0}$
can be chosen so that we have
$$
\frac{\pt {\cal D}}{\pt u}\Big|_{u=0} \ne 0\,.
$$
It must be pointed out that the above argument falls short of being a
rigorous proof: the function $M$ has
a global character, and in
particular it {\em might not} be independent of $\spm_2\Big|_{u=0}$ and
$\gamma_{3,1}\Big|_{u=0} $.
 (Nevertheless the above calculation shows
that no obvious miraculous cancellations occur, and we  find it completely
implausible that in, say, the vacuum case there exists some kind of
conspiracy between the functions $\spm_2\Big|_{u=0}$,
$\gamma_{3,1}\Big|_{u=0} $, $c$ and $M$ which leads to the {\em
identical} vanishing
of the $\theta$--integral of $G$.)

It is curious, and not entirely unexpected, that in the axisymmetric
polyhomogeneous setting there is a Newman--Penrose type quantity which
is again a constant of motion. Define
\be
\label{newlogconservedquantity}
{\cal Q} = \int_{S^2} \spm_{4,1} \sin^2 \theta\, d\mu_0\, .
\ee
As we show in Appendix \ref{npconstapp}, ${\cal Q}$ is
conserved by the evolution via the vacuum Einstein equations. It seems
clear to us that an analogous result will be true in the general case,
without assuming axisymmetry.

\section{Existence of Bondi coordinates}
\label{coordinates}
\newcommand{\mg}{\gamma}
\newcommand{\ap}{{\cal A}_{phg}}
Consider a metric $\mg$ defined on the set
\be
\label{BC.0}{\cal U}_{\hat R,\hat C_1,\hat C_2}\equiv\{\hat r \ge \hat R,\,
\hat u\in (\hat C_1,\hat C_2),\, \hat x^a\in S^2\},
\ee
and suppose that there exists $0<\epsilon < 1$ such that
\be
\label{BC.1.0}
\mbox{ for}\ \ (\hat x^\mu\hat x^\nu)=(\hat u\hat u),(\hat u\hat x^a),(\hat
r\hat x^a)\qquad
\mg_{\hat x^\mu\hat x^\nu}\in \ap,\qquad |\mg_{\hat x^\mu\hat x^\nu}   |
\le
\epsilon^{-1}\ ,
\ee
\be
\label{BC.1}
\mg_{\hat r\hat u}\in \ap,\qquad \epsilon \le \mg_{\hat r\hat u}\le
\epsilon^{-1}\ ,
\ee
\be
\label{BC.2}
\mg_{\hat r\hat r}\in \hat r^{-2}\ap,\qquad | \hat r^{2}\mg_{\hat r\hat r}   |
\le
\epsilon^{-1}\ ,
\ee
\be
\label{BC.4}
\mg_{\hat x^b\hat x^a}\in \hat r^{2}\ap,\qquad | \hat r^{-2}\mg_{\hat
x^b\hat x^a}   |
\le
\epsilon^{-1}\ .
\ee
Following Penrose \cite{Penrose} let $\hat x
:= \hat r^{-1}$, and set
\newcommand{\tg}{\tilde \gamma}
$$
\tg_{\mu\nu} \equiv \hat x^{2}\mg_{\mu\nu}
\ .
$$
{}From \eq{BC.0}--\eq{BC.4} it follows that the metric
$\tg_{\mu\nu}dx^\mu dx^\nu$
can be extended by continuity to a polyhomogeneous metric on the set
\be
\label{BC.5}
{\cal V}_{1/\hat R,\hat C_1,\hat C_2}\equiv\{0\le\hat x\le1/\hat R,\,
\hat u\in(\hat C_1,\hat C_2),\,\hat x^a\in S^2\}\
{}.
\ee
Actually the conditions \eq{BC.1}--\eq{BC.4} guarantee
only that
the appropriately rescaled functions $\tg_{\mu\nu}$ can be extended by
continuity to
the boundary, with the appropriately rescaled metric degenerating
perhaps at the boundary. We thus add the
supplementary restriction that $\tg_{ab}dx^a dx^b$ is {\em
non--degenerate up--to--boundary},
and that $\tg$ is  also {\em non-degenerate with signature
$(-+++)$ up--to--boundary}, in a sense which should be clear from what
is said below.
Define
$$
{\scri}= \{\hat x=0,\,
\hat u\in(\hat C_1,\hat C_2),\,\hat x^a\in S^2\}\ .
$$
Throughout this paper we shall suppose that ${\scri}$ is a null
hypersurface; as in the smooth case ({\em cf.\ e.g.\
}\cite{Wald}), in the polyhomogeneous
case this will  necessarily hold if $\gamma$ is vacuum.
Let $\bu$ be any smooth function on $\scri$,
and extend $\bu$ to a smooth
function defined in some neighbourhood of $\scri$ in
any way. Let $w_\mu dx^\mu$ be any smooth nowhere vanishing
one-form field defined in a
neighbourhood of $\scri$
such that $w_\mu X^\mu=0$ for all $X^\mu\in T\scri$; from the fact that
$\scri$ is null it is easily seen that
\be
(\tilde \gamma ^{\mu\nu}w_\mu w_\nu)\bigg|_{\scri}=0\ .
\label{null}
\ee
On $\scri$ consider the one-form field
\be
k_\mu\bigg|_{\scri}dx^\mu \equiv a \, w_\mu\bigg|_{\scri}
dx^\mu+d\bu
\label{BC.6}
\ee
with some function $a$; from (\ref{null}) it follows that the equation
\be
(\tg^{\mu\nu}k_\mu k_\nu)\bigg|_{\scri}
=(2 a \tg^{\mu\nu}w_\mu \bu_\nu + \tg^{\mu\nu}\bu_\mu \bu_\nu)\bigg|_{\scri}=0\
\label{BC.7}
\ee
$\displaystyle{(\bu_\mu\equiv\frac{\partial \bu}{\partial x^\mu})}$
will have a (unique) smooth solution $a|_\scri\,$ provided that
\be
\label{cond:solv}
(\tg^{\mu\nu}w_\mu \bu_\nu)\bigg|_{\scri}\qquad \mbox{ is bounded away
from zero,}
\ee
which we shall assume to hold. (Note that (\ref{cond:solv}) implies
that $k^\mu\equiv \tilde\gamma^{\mu\nu}k_\nu$ will be transverse to $\scri$.)
By Proposition \ref{P6} point 1 for every $(\bu,\hat x^a)\in\scri$
there exists a null geodesic $x^\mu(s,\bu,\hat x^a)$ such that
$dx^\mu/ds(0,\bu,\hat x^a)=k^\mu|_\scri$\ . There also exists a
diffeomorphism $\hat x^a\rightarrow \breve x^a(\hat x^b)$ such that in the
coordinates $(\bu,\breve x^a)$ we have at $\bu=0$
$$
\tg_{ab}\bigg|_{u=x=0}d\breve x^ad\breve x^b=\phi^2\breve
h_{ab}d\breve x^ad\breve x^b\equiv
\phi^2(d\theta^2+\sin^2\theta\, d\varphi^2)\ ,
$$
where $\phi\in C^\infty(S^2)$ is uniformly bounded away from zero.
Let $(u,s,x^a)$ be obtained by Lie dragging $(\bu,\breve x^a)$ along the
integral curves $\hat x^\mu(s,\bu,\hat x^a)$. Thus if we set $k^\mu\equiv
\partial \hat x^\mu/\partial s$, then
\be
 k^\mu u,_\mu=0, \qquad u(0,\bu,\breve x^a)=\bu\ ,
\label{BC.8}
\ee
\be
 k^\mu x^a,_\mu=0, \qquad x^a(0,\bu,\breve x^b)=\breve x^a\ .
\label{BC.9}
\ee
By point 2 of Proposition \ref{P6}
and by the implicit function theorem there exists a neighbourhood of
$\scri$ on which $(u,s,x^a)$ form a coordinate system.
As $k^\mu$ is tangent to null geodesics we have $\tg_{\mu\nu}k^\mu
k^\nu=0$, so that
$$
\tg(\frac{\partial}{\partial s},\frac{\partial}{\partial s})=0 \ .
$$
It follows that  in these coordinates $\tg$ takes the form
\be
\tg_{\mu\nu}d\hat x^\mu d\hat x^\nu=\tg_{uu}du^2+2\tg_{us}du\,ds
+\tg_{ab}(dx^a+U^adu)(dx^b+U^bdu)\ .
\label{BC.10}
\ee
Setting $\check r = 1/s$ the original metric $\mg$ takes the form
$$
\mg_{\mu\nu}d\hat x^\mu d\hat x^\nu= \hat r^2
\tg_{uu}du^2-2(\frac{\hat r}{\check
r})^2
\tg_{us}dud\check r+ \hat r^2 \tg_{ab}(dx^a+U^adu)(dx^b+U^bdu)\ ,
$$
where all functions in $\gamma_{\mu\nu}$ are polyhomogeneous, after
appropriate rescalings.
Finally define
\be
r\equiv
\left[\sin^2 \theta \, \det\left(\gamma^{\mu\nu}\frac{\partial
x^a}{\partial \hat x^\mu} \frac{\partial
x^b}{\partial \hat x^\nu} \right)\right]^{-1/4}\ .
\label{BC.10.1}
\ee
One easily finds
$$
\frac{\partial r
}{\partial\hat r }=\phi
+O(\frac{\log^N\check r}{\check r})
$$
for some $N$,
so that there exist constants $R,C_1,C_2$ such that $(r,u,x^a)$ as
constructed above form a coordinate system on ${\cal U}_{R,C_1,
 C_2}$, where ${\cal U}_{ R, C_1, C_2}$ is given
by \eq{BC.0} (in the coordinates $(r,u,x^a)$).
Going to this  coordinate system from \eq{BC.10.1} one concludes that
$$
\det
(\gamma_{ab})=r^4\sin^2\theta\ .
$$
In this way one obtains
a metric of the form \eq{(B5.1)}, which satisfies appropriate Bondi
requirements at $u=0$.
If one moreover assumes that $\gamma$ is vacuum, then the Einstein
equations imply
$$
\frac{\partial}{\partial u}\left(\lim_{r\rightarrow\infty}
\tg_{ab}\right) = 0\ ,
$$
so that we have
$$
\lim_{r\rightarrow\infty}
\tg_{ab}=d\theta^2+\sin^2\theta d\varphi^2
$$
for all $u\in(C_1,C_2)$. As discussed in Section \ref{analysis} we can
achieve
$$
\lim_{r\rightarrow\infty} U^a = 0
$$
by appropriately
propagating the coordinates $x^a$ away from the surface $u=0$.

The above construction shows that the Bondi coordinates above are
uniquely determined by the choice of a function $\bu = u\big|_\scri$\ .
Let us show that the freedom in the choice of $u\big|_\scri$\ can be
considerably reduced if we require
\be
\lim_{r\rightarrow\infty} \beta = 0.
\label{BC.11}
\ee
To achieve \eq{BC.11}, let  $\bar u\big|_\scri$\ be
given by
\be
\bar u(u,x^a)=\int_0^u e^{2H(u',x^a)}du'+\alpha(x^a)\ ,
 \qquad (H=\lim_{r\rightarrow\infty} \beta )\ ,
\label{BC.13.0}
\ee
where $\alpha\in C^\infty(S^2)$ is an arbitrary function. We can now
repeat the construction described above of a coordinate system
$(\bar s,\bau,\bar x^a)$ based upon  this function $\bar
u\big|_\scri$\ , obtaining again a metric of the form \eq{BC.10} in this
coordinate system.
 Let thus a vacuum metric $\gamma$ of the form (\ref{(B5.1)}) be
given, and set
$\tilde \gamma_{\mu\nu}=r^{-2}\gamma_{\mu\nu}$, $x=r^{-1}$,
$k^\mu=\tilde\gamma^{\mu\nu}\bar u,_\nu$; from the equality
$$
\tilde\gamma^{\mu\nu}\bigg|_\scri\partial_\mu\partial_\nu
=
-2e^{-2H}\partial_u\partial_x + \breve h^{ab}\partial_a\partial_b
$$
by construction of $k^\mu$ we obtain
\be
k^\mu\bigg|_\scri\partial_\mu=
-\partial
_x-e^{-2H}\frac{\partial\bau}{\partial x}\bigg|_\scri\partial_u
+{\cal D}^a\bau\bigg|_\scri\partial_a\ ,
\label{BC.13}
\ee
with
\be
\frac{\partial\bau}{\partial x}\bigg|_\scri
=\frac{1}{2}|{\cal D}\bau|_{\breve h}^2\bigg|_\scri\ ,
\label{BC.14}
\ee
where ${\cal D}$ denotes the covariant derivative of the metric
$\breve h$ on $S^2$. We also have
\be
\frac{\partial \bar x^a}{\partial u}\bigg|_\scri=0\ ,
\quad
\frac{\partial \bar x^a}{\partial x^b}\bigg|_\scri=\delta^a_b\ ,
\quad
\frac{\partial \bar x^a}{\partial x}\bigg|_\scri=
{\cal D}^a\bau\bigg|_\scri\ .
\label{BC.16}
\ee
{}From the ``barred'' equivalent of \eq{BC.10.1},
$$
\bar r = \left[ \sin^2 \theta \, \det\left(\gamma^{\mu\nu}\frac{\partial
\bar x^a}{\partial x^\mu} \frac{\partial
\bar x^b}{\partial x^\nu} \right)\right]^{-1/4}\ ,
$$
and from what has been said one easily finds
\be
\lim_{r\rightarrow\infty}\frac{\partial \bar r}{\partial r}
=1\ ,\qquad
\lim_{r\rightarrow\infty}\frac{\partial \bar r}{\partial u}
=\lim_{r\rightarrow\infty}\frac{\partial \bar r}
{\partial x^a}
=0\ .
\label{BC.17}
\ee
Equations  \eq{BC.14}--\eq{BC.17} show that in the coordinate system
$(\bar u,\bar r,\bar x^a)$ the metric takes the desired form with
\be
\lim_{r\rightarrow\infty} \beta = 0\ .
\label{BC.18}
\ee
It is worthwhile mentioning that the only freedom in the choice of
coordinates left at this stage is that of the function $\alpha$ in
\eq{BC.13.0}. Any two coordinate systems $(u_1,r_1,x^a_1)$ and
$(u_2,r_2,x^a_2)$ which satisfy our requirements will have the
property that
$$
\alpha\equiv (u_1-u_2)\bigg|_\scri
$$
is a function of $x^a$ only, and the coordinate system
$(u_2,r_2,x^a_2)$ is defined uniquely by $\alpha$ and by
$(u_1,r_1,x^a_1)$. It follows that for polyhomogeneous metrics the
asymptotic symmetry group is the BMS group, as in the smooth case.
\vspace{15pt}

 {\bf Acknowledgements.} Most of the work on this paper was performed
when P.T.C.\ was visiting the Centre for Mathematics and
its Applications of the ANU.
P.T.C.\ also wishes to acknowledge the friendly
hospitality of the Department of Mathematics of Universit\'e de Tours
during part of work on this paper. We  also thank Inge Frick,
Jan {\AA}man, James Skea and Anthony Hearn, the principal authors of the
software we used in this paper. We thank Prof. Sir Hermann
Bondi, F.R.S., for permission to publish this
paper as one of the series he initiated.

\appendix

\appsection{Conventions, function spaces}
\label{conventions}

We shall assume that all manifolds we discuss (which will have
dimension 2, 3 or 4) are paracompact, connected, Hausdorff and smooth.
The summation convention is used throughout this paper. Space--time is
as usual taken to be a Lorentzian 4--dimensional manifold with
signature  $(-,\,+,\,+,\,+)$ and
metric connection ${\Gamma^\lambda}_{\mu\nu}$ where Greek indices run
from 0 to 3; its Riemann curvature tensor is defined by
 $$
{R^\alpha}_{\beta\gamma\delta} =
\partial_\gamma{\Gamma^\alpha}_{\beta\delta}-
\partial_\delta{\Gamma^\alpha}_{\beta\gamma}+
{\Gamma^\alpha}_{\mu\gamma}{\Gamma^\mu}_{\beta\delta}-
{\Gamma^\alpha}_{\mu\delta}{\Gamma^\mu}_{\beta\gamma}
 $$
and the Ricci tensor and scalar by
$$R_{\beta\delta}={R^\alpha}_{\beta\alpha\delta},
 \quad R=g^{\mu\nu}R_{\mu\nu}.$$
With these choices the Einstein equations take the form
$$R_{\mu\nu}-\frac{1}{2}Rg_{\mu\nu} = \kappa T_{\mu\nu}\,,$$
with a positive constant $\kappa$.

In this paper we are considering asymptotic behaviour near null
infinity. We will use
$\bar M$ to denote a manifold with boundary so that
$M\equiv {\rm int}~ \bar M$ is a manifold
of dimension $n$ and $\pt M\equiv \pt \bar M$ is a manifold which
will be assumed to have a finite number of connected components $\pt
M_i$.
By an abuse of terminology $M$ will also be said to be a manifold with
boundary. As usual $T_pM$ will denote the tangent space to $M$ at $p$;
for $p$ in $\partial M$ one has the notion of ``half--tangent space at
$p$'' which is defined in a natural obvious way, and we shall still write
$T_pM$ for this space.

Throughout the paper $x$ will denote a defining function for $\pt
M$, {\em i.e.\ }a function satisfying $x\big|_{\pt M}=0$, $x\ge 0$,
$dx(p)\ne 0$ for $p\in{\pt M}$,
and the implication $x(p)=0\Rightarrow p\in\pt M$
holds.

We can always choose a finite number of coordinate
charts $\phi_j : {\cal O}_j \ra \R^{n,+}\equiv \{y\in \R^n : y^1 \ge 0\}$,
$j=1,\ldots,J$, covering a neighbourhood of $\pt M$ such that $y^1 = x$.
When referring to local coordinates we shall implicitly assume that $y^1 =
x$, and we shall use the letter $v$ to denote the coordinates
$y^2,\ldots,y^n$;
$$v^A = y^A\,, \quad A = 2,\ldots,n\,.$$
Thus $y=(x,v)\ .$
The standard Schwarz  multi--index notation is used
throughout; thus if
$\alpha=(\alpha_1,\ldots \alpha_n)$, then
$$ \pt^\alpha=\pt^\alpha_y=\pt^{\alpha_1}_{y^1}\cdots
\pt^{\alpha_n}_{y^n}=\pt^{\alpha_1}_{x} \pt^{\alpha_2}_{y^2}\cdots
\pt^{\alpha_n}_{y^n}=\pt^{\alpha_1}_{x} \pt^{\beta}_{v}\
,$$
where $\beta=(\alpha_2,\ldots,\alpha_n)$.

For
$k\in\N_0^\infty\equiv \N\cup\{0\}\cup\{\infty\}$ the spaces $C^k_{\rm
\lint}(M)$
are the spaces of functions $k$--times differentiable on $M$.
We have added the subscript
``\lint'' to emphasize
the fact  that a function in  $C^k_{\rm \lint}(M)$ need {\em not}
extend to the boundary of $M$ (in this respect the subscript ``\lint''
does not imply the same sense of ``local'' as the local coordinates we
have just defined);
similarly  for $k\ge 1$  even if the function itself extends by
continuity to $\partial M$  then its derivatives do not have to
extend, etc. We use the symbol  $C^k({\bar M})$
for  the Banach spaces of functions differentiable $k$-times on
$M$ such that $f$ and its derivatives up to order $k$
can be extended to {\em
continuous\/} functions on ${\bar M}$, and equipped with the supremum
norm.

Let $f_i$ be a sequence of functions in $C^\infty_{\rm \lint}(M)$ and let
$x_1>0$; we suppose that,
given $N\in \N$, there is a sequence
$s_{i,N}
{\renewcommand{\arraystretch}{0.13}
\begin{array}[t]{c}
 \longrightarrow \\ {\scriptscriptstyle
i\rightarrow \infty} \end{array}} \infty$
and some constants
$C_{i,N}$
such that for all $|\alpha| \le N$ and
for all $0<x\le x_1$,
$$
\big|\pt_y^\alpha f_i\big| \le C_{i,N} x^{s_{i,N}}\,,
$$
To express the notion of successive approximations good to
all powers of $x$ and for all derivatives
in a precise sense, we shall write
$$f \sim \sum_{i=0}^\infty\, f_i$$
if for every $n,m \in \N$ there exists $N \in \N$ and a constant $C(n,m)$
such that
for all $|\alpha| \le m$
and for $0 < x \le x_1$
$$
\bigg|\pt_y^\alpha \bigg(f - \sum_{i=0}^N\, f_i\bigg)\bigg| \le C{(n,m)}\,
x^n\,.
$$

Consider a sequence $\{N_{j}\}_{j=0}^\infty$, $N_{j} \in \N_0$. $f$ will
be said to be polyhomogeneous
if $f \in C^\infty_{\rm \lint}(M)$ and there exists a sequence of functions
$f_{jk} \in C^\infty(\bar M)$ such that
\begin{equation}
\label{sim}
f \sim
\sum_{j=0}^\infty\, \sum_{k=0}^{N_{j}}\,
f_{jk}\, x^{j}\, \log^k x\,.
\end{equation}
We write $f \in {\cal A}^{\{N_{j}\}}$, and define
${\cal A}_{phg}\equiv\cup_{\{N_{j}\}}{\cal A}^{\{N_{j}\}}$.

 {\bf Remark:} As formulated here, the $f_{ij}$ may depend upon
$x$. To avoid this would require the introduction of local coordinates
near the boundary, a specialization which we do not yet wish to make.
However, once we do fix a coordinate system, then we can Taylor expand
each $f_{ij}$ with respect to $x$ to any finite order, so each
$f_{ij}$ has a polyhomogeneous expansion with no log terms, and obtain
an expansion for $f$ with some (other) functions $f_{ij}$ which depend
only upon the coordinates $v$.

A function $f(r,v)$ defined on an open set of the form ${\cal O}=\{(r,v): r\in
(r_0,\infty), v\in Q\}$ for some suitable set $Q$
will be said to be in $C^k(\bar {\cal O})$,
$0\le k\le \infty$, if
$f(1/x,v)\in C^k(\bar {\cal U})$, where ${\cal U}=\{(x,v): x\in (0,1/r_0),
v\in Q\}$. Similarly $f$ will be said to be polyhomogeneous on ${\cal O}$
if $f(1/x,v)$ is polyhomogeneous near $x=0$ on ${\cal U}$.

Let $F$ be a function space over $M$.
A tensor field $X = ({X^\alpha}_\beta)$,
where $\alpha, \beta$ are some multi--indices, $|\alpha| = r$, $|\beta|
= s$, will be said to belong to $F$
if  in local coordinates as described at the beginning of this section the
components ${X^\alpha}_\beta$ of $X$ are in $F$.

Let $F$ be a function space. We shall write that $f \in r^\alpha\, \log^\beta
r\  F$ if $r^{-\alpha}\, \log^{-\beta} r \ f \in F$.

Let $F_1, F_2$ be function spaces. We shall write that $f \in F_1 + F_2$ if
there exist $f_a \in F_a$, $a = 1, 2$, such that $f = f_1 + f_2$.

The following observations are useful when proving Proposition \ref{P1}:

\begin{Proposition}:
\label{PA1}
\begin{eqnarray}
f \in r^\alpha\, \log^\beta r\, {\cal A}_{phg}, \quad g \in r^\mu\,
\log^\nu r\, {\cal
A}_{phg} &\Longrightarrow
&fg \in r^{\alpha+\mu}\, \log^{\beta+\nu} r\, {\cal
A}_{phg}\ ;
\label{(A2.1)}
\\
F \in C^\infty(\R),\quad f \in C^\infty(\bnh) + \frac{{\cal A}_{phg}}{r}
& \Longrightarrow & F(f) \in {\cal A}_{phg}\ ;
\label{(A2.2)}
\\
i \in \Z,\quad
f \in r^i\, {\cal A}_{phg} &\Longrightarrow &\forall j \in
\N,\quad \pt_r^j\, f \in r^{i-j}\, {\cal A}_{phg}\ ,
\label{(A2.3)}
\\
 & &  \forall\; \alpha,\quad \pt_v^\alpha\, f \in r^i{\cal
A}_{phg}\ ;
\label{(A2.4)}
\\
i \in \Z,\quad  f \in r^{i}\, {\cal A}_{phg} &\Longrightarrow&
\int_{R}^r f \in r^{i+1}\, {\cal A}_{phg}\  .
\label{(A2.5)}
\end{eqnarray}
\end{Proposition}

\smallskip
\noindent{\bf Proof}: (\ref{(A2.1)}) is easily proved starting with the
following elementary observations: $f, g \in {\cal A}_{phg}
\Longrightarrow f + g \in {\cal A}_{phg}$; $f \in r^\alpha\, \log^\beta r\,
C^\infty(\partial M)$ (note the Remark above), $g\in {\cal A}_{phg}
\Longrightarrow fg \in r^\alpha\,
\log^\beta r {\cal A}_{phg}$. (\ref{(A2.2)}) follows similarly from
(\ref{(A2.1)}) by Taylor expanding $F$. (\ref{(A2.3)}) and (\ref{(A2.4)})
are elementary. (\ref{(A2.5)}) follows by linearity from
$$
\int r^i \log^j r\  dr = \left\{
{\renewcommand{\arraystretch}{2}
\begin{array}{ll}
\sum_{k=0}^j C_{ijk} r^{i+1} \log^kr , &\mbox{$i \ne -1$} , \\
\displaystyle{{{\log^{j+1} r} \over {j+1}}} , &\mbox{$i = -1$} ,
\end{array}} \right.
$$
for some coefficients $C_{ijk}$.

\appsection{Geodesics in polyhomogeneous metrics}
\label{geodesics}
\begin{Proposition}:
\label{P6}
Let $g$ be a polyhomogeneous metric (Lorentzian or Riemannian) on a
manifold with boundary $M$, $g \in {\cal A}_{phg} \cap C^0(\bar M)$.
\begin{enumerate}
\item
For any $p \in \pt M$, $k \in T_p \bar M$
there exists $\epsilon > 0$ and a
unique geodesic $\Gamma_p(s)$, $s \in [0, \epsilon)$, satisfying
\begin{equation}
\Gamma_p(0) = p\,,\qquad {\dot \Gamma}_p(0) = k\,.\label{(geo.0)}
\end{equation}
If we write $\Gamma_p = \{y^\mu(s)\}$, then $y^\mu(s)$ are
polyhomogeneous functions of $s$.
\item
If $\pt M \ni p \ra k_p$ is a smooth field, $k \in C^\infty(\pt M)$, then the
functions $y^\mu(s, v)$ are polyhomogeneous.
\end{enumerate}
\end{Proposition}

\noindent{\bf Remark}: For a polyhomogeneous metric we have $\pt g
\sim \log^{N_1} x$ near $\pt M$, so that standard results about
geodesics do not apply.

\smallskip
\noindent{\bf Proof}: Let us define
$$\psi^\mu(s) = y^\mu(s) - sk^\mu - y_0^\mu\,,$$
we thus have
\begin{equation}
\frac{d^2 \psi^\mu}{ds^2} = F^\mu(\psi, {\dot \psi}, s)\,,\label{(geo.1)}
\end{equation}
with
\begin{equation}
F^\mu(\psi, \chi, s) = \Gamma_{\alpha\beta}^\mu (y_0^\nu + sk^\nu +
\psi^\nu)\, \chi^\alpha\, \chi^\beta\,.\label{(geo.2)}
\end{equation}
Let $\alpha \in (0, \frac{1}{2})$ and let $\epsilon > 0$ be a number to be
determined later. Consider the space
$$X_\epsilon = \big\{\psi^\mu, \chi^\mu \in C([0, \epsilon])\big\}$$
with the norm
$$\|(\psi, \chi)\big\|_{X_\epsilon} = \sup_{s\in[0,\epsilon]} |s^{-\alpha}
\chi^\mu(s)| + \sup_{s\in[0,\epsilon]} |s^{-\alpha-1} \psi^\mu(s)|\,.$$
Let $T : X_\epsilon \ra X_\epsilon$ be the map
\begin{equation}
X_\epsilon \ni (\psi^\mu, \chi^\mu) \longrightarrow T [\psi, \chi] =
\left(\int_0^s \chi^\mu(s) ds, \int_0^s F^\mu(\psi, \chi, s)\,
ds\right)\,,\label{(geo.3)}
\end{equation}
$F^\mu$ given by (\ref{(geo.2)}). Clearly a fixed point of $T$ satisfies
(\ref{(geo.0)})--(\ref{(geo.1)}). From the estimates
\begin{eqnarray}
|\Gamma| + |\pt_v \Gamma| &\le &C \log^N x\,,\label{(geo.4)}\\
|\pt_x \Gamma| &\le &C x^{-1} \log^N x\,,\label{(geo.5)}
\end{eqnarray}
for some $N$, it is easily seen that one can choose a constant $K$ and an
$\epsilon > 0$ such that $T$ is a contraction mapping from a ball around
$(0, 0)$ in $X_\epsilon$ of radius $K$ into itself, and the result follows by
the contraction mapping principle. Once the solution is known to exist,
polyhomogeneity immediately follows from the equation
$$(\psi, \chi) = T [\psi, \chi]\,.
$$
If $k^\mu(v)$ is a smooth function of $v \in \pt M$, then by considering
the equations satisfied by $\pt_v^\alpha y^\mu$ (note that for a
polyhomogeneous metric $\pt_v^\alpha \Gamma$ satisfies the same
estimates (\ref{(geo.4)})--(\ref{(geo.5)}) as $\Gamma$ itself) one obtains
the result by an argument similar to the one above.\hfill$\Box$

Let us finally point out that part 1 of Proposition \ref{P6} is still
true under the rather weaker hypotheses
\be
\label{weaker}
|g^{\mu\nu}| +|g_{\mu\nu}| + x^{1-\beta}|\partial_\sigma g_{\mu\nu}| +
x^{2-\beta}|\partial _\sigma\partial_\rho g_{\mu\nu}|  \le C
\ ,
\ee
for some constant C, with any $\beta>0$. The same proof as above goes
through,
except that the exponent $\alpha$ in the norm $\|(\psi,
\chi)\big\|_{X_\epsilon}$ has to be chosen to lie in $(0,\beta)$.
Gauss coordinates can be constructed for metrics satisfying
(\ref{weaker})
provided that one moreover has
\be
\label{weaker1}
x^{1-\beta}|\partial_A\partial_\rho g_{\mu\nu}|
+x^{2-\beta}|\partial_A\partial_\rho\partial_\sigma g_{\mu\nu}|
\le C
\ ,
\ee
where $\partial_A$ are derivatives in directions tangent to
$\partial M$, while $\partial_\rho$, etc., denotes all partial derivatives.

\appsection{The Einstein field equations for the metric (2.1)}
\label{vdbeqs}
Using {\sc sheep} we have derived the Einstein equations for a metric of the
form (\ref{(B5.1)}). The metric coefficients have been further
parametrized as in Proposition \ref{P4} with the small changes that,
to aid comparison with \cite{vdB}, $U^\theta$ is written as $U$ and
$U^\phi$ is written as   $W \cosec \theta$. Derivatives with
respect to the coordinates, which are numbered by
$$(x_0,\,x_1,\,x_2,\,x_3) = (u,\, r,\, \theta,\, \phi)$$
are indicated by subscripts.
The equations one obtains
coincide with those of
the Appendix in \cite{vdB}, except for some misprints listed below.
If all terms were
moved to the left hand sides, the left sides would be
$rR_{11}/4$ in \eq{vdb1},
$2r^2R_{12} $ in \eq{vdb2}, $2r^2R_{13} \cosec \theta$  in \eq{vdb3},
$-e^{2\beta}(h^{ab}R_{ab})/2$ in \eq{vdb4},
$e^{2\beta}(e^{-2\gamma}R_{22} -e^{2\gamma} \sin^{-2} \theta
R_{33})/4r$  in \eq{vdb5} and
$e^{2\beta}(R_{23}-(h^{ab}R_{ab})h_{23}/2)/(2r \cosh 2\delta \sin
\theta)$ in \eq{vdb6}.

The misprints in the corresponding equations of
\cite{vdB} are:
\begin{enumerate}
\item
on p.\ 121, at the end of the second line of the second equation,
a right parenthesis is missing;
\item
on p.\ 122, in the fourth line of the equation the signs
of the terms which contain $\beta_{23}$ and $\beta_{2}\beta_{3}$ should
be reversed, and
\item
in the fifth line of the same equation the factor $r$ should be
replaced by $r^3$.
\end{enumerate}

\be
\label{vdb1}
  \beta_{ 1}
  = \frac{r}{ 2} (\gamma_{ 1}^{2} \cosh^{2} 2\delta  +  \delta_{1}^{2})
\ee
\begin{eqnarray}
\lefteqn
{\left(r^4 e^{-2\beta}(e^{2\gamma}U_1 \cosh 2\delta +W_1 \sinh
2\delta)\right)_1}&&\nn \\
\quad & = &
 2 r^{2}
  \Big( \beta_{ 1 2 } +2 \delta_{ 1} \delta_{ 2 }
	         -2 r^{-1} \beta_{ 2 }
    - 4 \gamma_{ 1} \delta_{ 2} \cosh 2 \delta \sinh 2 \delta  \nn \\
&&    - \left( \gamma_{1 2 } -2 \gamma_{ 1} \gamma_{ 2 }
               + 2  \gamma_{ 1} \cot \theta \right)
      \cosh^{2} 2\delta \Big)
    \nn \\
&& + 2 r^{2} e^{2 \gamma } \cosec \theta
   \Big( - \delta_{ 1 3} -2 \delta_{ 1} \gamma_{ 3 }
    + \left( \gamma_{ 1 3}  +2  \gamma_{ 1} \gamma_{ 3} \right)
    \cosh 2 \delta \sinh 2 \delta  \nn \\
&&
   \qquad\qquad\qquad\qquad +2  \gamma_{ 1} \delta_{ 3 } ( 1+2 \sinh^2
2\delta ) \Big)
\label{vdb2} \\
&&\nn\\
\lefteqn{
\left(r^4 e^{-2\beta}(U_1 \sinh 2\delta +e^{-2\gamma} W_1 \cosh
2\delta)\right)_1}&&\nn \\
\quad & =&
 2 r^{2} e^{- 2 \gamma }
  \Big( - \delta_{ 1 2 } +2 \delta_{ 1} \gamma_{ 2 }
        - 2 \delta_{ 1} \cot \theta
    -\left( \gamma_{ 1 2 } -2 \gamma_{ 1} \gamma_{ 2 }
         + 2 \gamma_{ 1} \cot \theta  \right)
    \cosh 2 \delta \sinh 2 \delta  \nn \\
&&
 - 2  \gamma_{ 1} \delta_{ 2 } ( 1 + 2 \sinh^{2} 2 \delta )
   \Big)
  +2 r^2 \cosec \theta
    \left( \beta_{ 1 3} +2  \delta_{ 1} \delta_{ 3} -2 r^{-1} \beta_{ 3 }
      \right. \nn \\
&&   +4 \gamma_{ 1} \delta_{ 3} \cosh 2 \delta \sinh 2 \delta
    +\left( \gamma_{ 1 3} +2 \gamma_{ 1 } \gamma_{ 3} \right)
         \cosh^{2} 2 \delta \Big)
\label{vdb3} \\
&&\nn \\
V_1 &=& 2 e^{2 \beta} \cosec \theta
  \Big( (
     \beta_{ 2 3 } + \beta_{ 2 } \beta_{ 3 }
       +2 \delta_{ 2 } \delta_{ 3 } ) \sinh 2 \delta
   +( \delta_{ 2 3 } + \delta_{ 3} \cot \theta
    + \delta_{ 2 } \gamma_{ 3 }  - \gamma_{ 2 } \delta_{ 3 }
     \nn \\
&&
    +  \delta_{ 2 } \beta_{ 3 }
    + \beta_{ 2 } \delta_{ 3 } ) \cosh 2\delta \Big)
  - e^{2 \beta -2 \gamma } \Big( ( \beta_{ 22}
    + \beta_{ 2 }^{2} + \beta_{ 2 } \cot \theta
    +2 \gamma_{ 2 }^{2} +2 \delta_{ 2 }^{2} -1   \nn \\
&& \left.
    - \gamma_{ 22} -3 \gamma_{ 2 } \cot \theta
    -2 \beta_{ 2 } \gamma_{ 2 } \right) \cosh 2 \delta
   +\left( \delta_{ 22} + 3 \delta_{ 2 } \cot \theta
    +2 \beta_{ 2 } \delta_{ 2 }
    -4 \gamma_{ 2 } \delta_{ 2 } \right)
   \sinh 2 \delta  \Big) \nn \\
&& - e^{2 \beta +2 \gamma} \cosec^2 \theta
     \left( ( \beta_{ 33} + \beta_{ 3 }^{2}
      +2 \gamma_{ 3 }^{2} +2 \delta_{ 3 }^{2}
      + \gamma_{ 33} +2 \beta_{ 3 } \gamma_{ 3 } )\cosh 2 \delta
     +\left( \delta_{ 33} +2 \beta_{ 3 } \delta_{ 3 } \right. \right. \nn \\
&&     +4 \gamma_{ 3 } \delta_{ 3 } )
       \sinh 2\delta \Big)
   - \frac{1}{ 4}    r^{4} e^{- 2 \beta }
     \left( ( e^{2 \gamma } U_1^{2}
            + e^{- 2 \gamma } W_1^{2} ) \cosh 2 \delta
       + 2 U_1 W_1 \sinh 2 \delta \right) \nn \\
&&  + \frac{1}{ 2}  r \left( r U_{ 1 2 }  + r U_1 \cot \theta
             +4 U_{ 2 }  +4 U \cot \theta \right)
     + \frac{1}{ 2}  r^2\cosec \theta
         \left(  W_{ 1 3}  +4  W_3 \right)
\label{vdb4} \\
&&\nn \\
\lefteqn{
(r\gamma)_{01} \cosh 2\delta +2r(\gamma_0 \delta_1 +\delta_0 \gamma_1)
 \sinh 2\delta } \nn \\
\quad &=&
     \frac{1}{ 2} ( \gamma_{ 1} V_1 + \gamma_{ 11} V
          +  r^{-1} \gamma_{ 1} V ) \cosh 2 \delta
       +2 \gamma_{ 1} \delta_{ 1} V \sinh 2 \delta
    + \frac{1}{ 8}  r^{3} e^{-2\beta} ( e^{2 \gamma } U_1^{2}
         - e^{- 2 \gamma } W_1^{2} ) \nn \\
&& + \frac{1}{ 2} r^{-1} e^{2 \beta -2 \gamma } (
      \beta_{ 22} +  \beta_{ 2 }^{2} -  \beta_{ 2 } \cot \theta )
   - \frac{1}{ 2} r^{-1} e^{2 \beta +2 \gamma } (
     \beta_{ 33} + \beta_{ 3 }^{2} ) \cosec^2 \theta \nn \\
&&  + r^{-1} e^{2 \beta } \left( \beta_{ 2 } \delta_{ 3 }
    - \beta_{ 3 } \delta_{ 2 } \right) \cosec \theta
    + \frac{1}{ 4} r e^{2 \gamma }  \cosec \theta
       \Big( ( U_{ 1 3} + 2 r^{-1} U_{ 3 } )
      \sinh 2 \delta  \nn \\
&&     + 4 \delta_{ 1} U_{ 3 } \cosh 2 \delta \Big)
    - \frac{1}{ 4}r e^{- 2 \gamma } \Big( \left( W_{ 1 2 }-W_1 \cot
      \theta   \right) \sinh 2 \delta
    +  2 r^{-1} \left( W_{ 2 } -W \cot \theta   \right) \sinh 2 \delta
     \nn \\
&&
    +4 \delta_{ 1} \left( W_{ 2 } - W \cot \theta \right) \cosh 2
     \delta  \Big)
    - \frac{1}{ 4} r \left( U_{ 1 2 } + 2 r^{-1} U_{ 2 }
         - U_1 \cot \theta
    -  2 r^{-1}  U \cot \theta  \right. \nn \\
&& \left. + 4 r^{-1} \gamma_{ 2 } U + 4 \gamma_{ 1 2 } U
    +  2 \gamma_{ 2 } U_1 + 2 \gamma_{ 1} U_{ 2 }
    +  2 \gamma_{ 1} U  \cot \theta \right) \cosh 2 \delta \nn \\
&&  -r \left( \delta_{ 1} U_{ 2 } +2 \gamma_{ 1} \delta_{ 2 } U
    + 2 \delta_{ 1} \gamma_{ 2 } U
    -  \delta_{ 1} U \cot \theta \right) \sinh 2 \delta \nn \\
&& + \frac{1}{ 4} r \cosec \theta (  W_{ 1 3}
        + 2 r^{-1} W_{ 3 } - 4 r^{-1} \gamma_{ 3 } W
    - 4 \gamma_{ 1 3} W
    - 2 \gamma_{ 3 } W_1
    - 2 \gamma_{ 1} W_{ 3 } ) \cosh 2 \delta \nn \\
&& + r \cosec \theta \left(
    \delta_{ 1} W_{ 3 }
    -2 \delta_{ 1} \gamma_{ 3 }  W
    -2  \gamma_{ 1} \delta_{ 3 } W \right) \sinh 2 \delta
\label{vdb5} \\
&&\nn \\
\lefteqn{
(r\delta)_{01} -2r\gamma_0\gamma_1 \sinh 2 \delta \cosh 2\delta} \nn \\
\quad & =&
  \frac{1}{ 2} ( \delta_{ 1} V_1 +  \delta_{ 11} V
  +  r^{-1} \delta_{ 1} V
   - 2  \gamma_{ 1}^2  V  \cosh 2 \delta \sinh 2 \delta  ) \nn \\
&&  - \frac{1}{ 2}  r^{-1} e^{ 2 \beta  -2 \gamma }
     ( \beta_{ 22} +  \beta_{ 2 }^{2}
       -  \beta_{ 2} \cot \theta ) \sinh 2 \delta
    - \frac{1}{ 2} r^{-1} e^{2 \beta +2 \gamma }  \cosec^2 \theta
        ( \beta_{ 33} + \beta_{ 3 }^{2} ) \sinh 2 \delta
    \nn \\
&&  - r^{-1} e^{ 2 \beta } \cosec \theta
  \left( - \beta_{ 2 3 }
    -  \beta_{ 2 } \beta_{ 3 }
    + \beta_{ 3 } \cot \theta
    + \beta_{ 2 } \gamma_{ 3 }
    - \gamma_{ 2 } \beta_{ 3 }  \right)  \cosh 2 \delta \nn \\
&&    + \frac{1}{ 8}  r^{3} e^{- 2 \beta} \left(
       ( e^{2 \gamma } U_1^{2}
    +  e^{- 2 \gamma} W_1^{2}  ) \sinh 2 \delta
    + 2 U_1 W_1 \cosh 2 \delta  \right) \nn \\
&&   - \frac{1}{ 2} r \Big(  2 \delta_{ 1 2 } U  + 2 r^{-1} \delta_{ 2 } U
        + \delta_{ 1} U_{ 2 }  + \delta_{ 2 } U_1
        + \delta_{ 1}  U \cot \theta
    - 2 ( \gamma_{ 1} U_{ 2 } -  \gamma_{ 1} U \cot \theta
         +4   \gamma_{ 1} \gamma_{ 2 } U )  \nn \\
&&  \quad \times \cosh 2 \delta \sinh 2 \delta \Big)
  - \frac{1}{ 2} r \cosec \theta \left(
  2 \delta_{ 1 3} W + 2 r^{-1} \delta_{ 3 } W + \delta_{1} W_{ 3 }
  + \delta_{ 3 } W_1 \right. \nn \\
&&
    + 2\left( \gamma_{ 1 } W_{ 3 } - 2 \gamma_{ 1 } \gamma_{ 3 } W \right)
   \cosh 2 \delta \sinh 2 \delta \Big)
  - \frac{1}{ 4}  r e^{- 2 \gamma } \Big(
     W_{ 1 2 } - W_1 \cot \theta
                                 \nn \\
&& \left.
  + 2 r^{-1} \left( W_{ 2 } - W \cot \theta  \right)
  -4 \gamma_{ 1} \left(W_{ 2 } - W \cot \theta  \right) \cosh^2 2
      \delta \right)
  - \frac{1}{ 4}  r e^{2 \gamma }  \cosec \theta \nn \\
&& \quad \times ( U_{ 1 3} +2 r^{-1} U_{ 3 }
     +4 \gamma_{ 1} U_{ 3 } \cosh^2 2 \delta )
\label{vdb6}
\end{eqnarray}

\appsection{Expansion in the axisymmetric case}
\label{npconstapp}

The following formulae give the expansion of the axisymmetric case
discussed at the end of section 3. All coefficients, except the last
one in each quantity ({\em i.e.\/} $\gamma_4$, $\beta_5$, $U_5$ and $V_3$) are
considered to be functions of $u$ and $\theta$ only: $\gamma_4$, $\beta_5$,
$U_5$ and $V_3$ are written as functions of $u$, $\theta$ and $r$ so
that the consistency of the approximation can be checked by looking at
the first neglected terms in the equations of appendix \ref{vdbeqs}.
The exponentials of $\beta$ and $\gamma$ are expanded to order $r^{-4}$.

As a result of the arguments of section 2, we know we can expand
$\gamma$ as
$$
\gamma = c(v)r^{ -1} +\gamma_{2}(v)r^{ -2} +\gamma_{3,1}(v)r^{ -3}\log r
 +\gamma_{3}(v)r^{-3} +\gamma_{4}(r,v)r^{ -4}\ ,
$$
where we use $(v)$ to denote $(u,x^a)$, and
the dependence of $\gamma_4$ on $r$ allows for $\log r$ terms there.
To avoid confusion of subscripts in what follows we use the form
$f_{,x}$ for the
partial derivatives, where $x$ is a variable name, not a number.

Substitution in (\ref{vdb1}) yields
$$
\beta = -\frac{1}{4}c^{2}r^{ -2}
-\frac{2}{3}c\gamma_{2}r^{ -3}
-\frac{3}{4}c\gamma_{3,1}r^{ -4}\log r
 + \left( \frac{1}{16}c\gamma_{3,1}
-\frac{3}{4}c\gamma_{3}
-\frac{1}{2}\gamma_{2}^{2} \right)r^{ -4}
+\beta_5 r^{-5} \nn
$$

Putting this into (\ref{vdb2}) gives
\begin{eqnarray}
U &= &
- \left( 2c\cot \theta
  +c_{,\theta} \right) r^{ -2}
- \left( \frac{8}{3}\gamma_{2} \cot \theta
   +\frac{4}{3} \gamma_{2,\theta} \right) r^{-3}\log r
+r^{ -3}U_{3} \nn \\
&& +\left( 2c \gamma_{2,\theta}
  +4c\gamma_{2}\cot \theta
  +\frac{3}{2} \gamma_{3,1,\theta}
  +3\gamma_{3,1}\cot \theta \right)  r^{ -4} \log r  \nn \\
&& + \left( \frac{5}{2}c^{3}\cot \theta
  +\frac{5}{4}c^{2}c_{,\theta}
  +\frac{5}{3}c\gamma_{2}\cot \theta
  -\frac{3}{2}cU_{3}
  +\frac{1}{2}c\gamma_{2,\theta}
  +\frac{2}{3}\gamma_{2}c_{,\theta} \right. \nn \\
&& \left.
  +\frac{11}{4}\gamma_{3,1}\cot \theta
  +3\gamma_{3}\cot \theta
  +\frac{11}{8}\gamma_{3,1,\theta}
  +\frac{3}{2}\gamma_{3,\theta} \right) r^{ -4}
+r^{ -5}U_5 \nn
\end{eqnarray}
where $U_3$ is as yet arbitrary. Its $u$-derivative is given by the
$R_{02}$ Einstein equation, and on substituting van der Burg's form for
the coefficient
$$
U_{3} = 4c^{2}\cot \theta
+3cc_{,\theta}
+2N$$
we find agreement with his equation for $N_{,u}$.

Next we use (\ref{vdb4}) to obtain
\begin{eqnarray}
V &=&
r -2M
+ \left( 2 \cot \theta \gamma_{2,\theta}
 -\frac{4}{3}\gamma_{2}
 +\frac{2}{3} \gamma_{2,\theta\theta} \right) r^{ -1}\log r \nn \\
&&+ \left( \gamma_{2,\theta} \cot \theta
  +\frac{5}{2}(c_{,\theta})^{2}
  +\frac{1}{3}\gamma_{2,\theta\theta}
  -\frac{1}{2}U_{3,\theta}
  +4c^{2}\cot^{2} \theta
  -\frac{3}{2}c^{2}
  -\frac{2}{3}\gamma_{2}  \right. \nn \\
&& \left. \quad  +\frac{3}{2}cc_{,\theta\theta}
  +\frac{19}{2}c c_{,\theta} \cot \theta
  -\frac{1}{2} U_{3}\cot \theta \right)r^{-1} \nn \\
&& + \left( 8c\gamma_{2}\cot^{2} \theta
  +4c \gamma_{2,\theta} \cot \theta
  +4\gamma_{2} c_{,\theta} \cot \theta
  +2 c_{,\theta}\gamma_{2,\theta} \right. \nn \\
&& \left. \quad  -\frac{1}{2} \gamma_{3,1,\theta\theta}
  -\frac{3}{2} \gamma_{3,1,\theta} \cot \theta
  +\gamma_{3,1} \right) r^{ -2}\log r  \nn \\
&&
+ \left( 4c^{3}\cot^{2} \theta
  +\frac{1}{6}c^{3}
  +\frac{9}{4}c^{2} c_{,\theta} \cot \theta
  -\frac{1}{4}c^{2}c_{,\theta\theta}
  +\frac{11}{6}c \gamma_{2,\theta}\cot \theta
   +\frac{1}{6}c\gamma_{2,\theta\theta}
  \right. \nn \\
&& \left.  \quad  +\frac{7}{3}\gamma_{2} c_{,\theta} \cot \theta
  +\frac{2}{3}\gamma_{2}c_{,\theta\theta}
  +\frac{7}{6}c_{,\theta}\gamma_{2,\theta}
  +\frac{4}{3}c\gamma_{2}\cot^{2} \theta
  -\frac{1}{3}c\gamma_{2}
  -3cU_{3}\cot \theta
  -\frac{3}{2}U_{3}c_{,\theta}
   \right. \nn \\
&& \left.
  -\frac{5}{8}\gamma_{3,1,\theta\theta}
  -\frac{15}{8} \gamma_{3,1,\theta} \cot \theta
  +\frac{5}{4}\gamma_{3,1}
  -\frac{1}{2}\gamma_{3,\theta\theta}
  -\frac{3}{2} \gamma_{3,\theta} \cot \theta
  +\gamma_{3} \right) r^{ -2} \nn \\
&&  +r^{ -3}V_3 \nn
\end{eqnarray}
where $M$ is as yet undetermined, but will have a $u$-derivative given
by the $R_{00}$ equation as
$$M_{,u} = \frac{3}{2}\cot \theta c_{,u\theta}-{c_{,u}}^2-{c_{,u}}
  +\frac{1}{2}c_{,u\theta\theta}. $$

On substituting all these expansions into the equation (\ref{vdb5}) we
find that $c_{,u}$ is undetermined, and $\gamma_{2,u} = 0$ as
expected. $\gamma_{3,1,u}$ is then $u$-independent so we can integrate
with respect to $u$ and get
$$
\gamma_{3,1} =  \gamma_{3,1}^{0}
+\left( \frac{1}{6}\gamma_{2,\theta\theta}
+\frac{1}{6} \gamma_{2,\theta}\cot \theta
-\frac{2}{3}\gamma_{2}\cot^{2} \theta
-\frac{1}{3}\gamma_{2} \right) u
$$
where $\gamma_{3,1}^{0}$ is
a freely specifiable function of $\theta$.

The next term in (\ref{vdb5}) gives the $u$-derivative of $\gamma_3$ as
\begin{eqnarray}
\gamma_{3,u} &=  &
\frac{3}{8}cc_{,\theta\theta}
+\frac{5}{8}c c_{,\theta} \cot \theta
+\frac{3}{8}(c_{,\theta})^{2}
-c^{2}\cot^{2} \theta
-\frac{1}{2}c^{2}
+\frac{1}{2}cM  \nn \\
&& -\frac{1}{12}\gamma_{2,\theta\theta}
-\frac{1}{12} \gamma_{2,\theta} \cot \theta
+\frac{1}{3}\gamma_{2}\cot^{2} \theta
-\frac{1}{3}\gamma_{2}
-\frac{1}{8}U_{3,\theta}
+\frac{1}{8}U_{3}\cot \theta \nn
\end{eqnarray}
and we cannot integrate this explicitly since we do not know the $u$
dependence of $c$ and hence of $M$ and $U_3$. We can check, using
$$
\gamma_{3} = C -\frac{1}{6}c^{3} $$
to get van der Burg's form of the coefficient,
that our result agrees with his when $\gamma_2 = 0$.

Finally we can solve the next order in (\ref{vdb5}) for $\gamma_{4,u}$ and
find that the result contains $\log r$ terms but no higher powers of
$\log r$, and that
$$
\gamma_{4,u} \sin^3 \theta = G + F_{,\theta}
$$
where $G$ is as in \eq{timederivative stuff} and
\begin{eqnarray}
F &=&  \Bigg(
 \left( -\frac{1}{24} \gamma_{2,\theta\theta\theta}\sin^{3} \theta
+\frac{1}{24} \gamma_{2,\theta\theta}\cos \theta \sin^{2} \theta
+\frac{7}{24}\gamma_{2,\theta}\cos^{2} \theta \sin \theta
\right. \nn \\
&& \left. \left. \quad \quad
+\frac{1}{8}\gamma_{2,\theta}\sin^{3} \theta
-\frac{2}{3}\gamma_{2}\cos^{3} \theta
-\frac{1}{2}\gamma_{2}\cos \theta \sin^{2} \theta
 \right) u \right.\nn \\
&&  \quad
-\frac{2}{3}c\gamma_{2}\cos \theta \sin^{2} \theta
-\frac{1}{3}c \gamma_{2,\theta} \sin^{3} \theta
-\frac{1}{4} (\gamma_{3,1}^{0})_{,\theta} \sin^{3} \theta
+\frac{1}{2}\gamma_{3,1}^{0}\cos \theta \sin^{2} \theta
 \Bigg)
\log r \nn \\
&&
+ \left(
-\frac{1}{32} \gamma_{2,\theta\theta\theta} \sin^{3} \theta
+\frac{1}{32}\gamma_{2,\theta\theta}\cos \theta \sin^{2} \theta
+\frac{7}{32} \gamma_{2,\theta}\cos^{2} \theta \sin \theta \right. \nn \\
&& \left. \quad +\frac{29}{96} \gamma_{2,\theta}\sin^{3} \theta
-\frac{1}{2}\gamma_{2}\cos^{3} \theta
-\frac{19}{24}\gamma_{2}\cos \theta \sin^{2} \theta
\right) u\nn \\
&&
-\frac{1}{4} \gamma_{3,\theta} \sin^{3} \theta
+\frac{1}{2}\gamma_{3}\cos \theta \sin^{2} \theta
+\frac{2}{3}\gamma_{2} c_{,\theta}\sin^{3} \theta
+\frac{1}{12}c\gamma_{2,\theta}\sin^{3} \theta
+\frac{5}{6}c\gamma_{2}\cos \theta \sin^{2} \theta  \nn \\
&& -\frac{11}{12}c^{3}\cos \theta \sin^{2} \theta
-\frac{7}{8}c^{2} c_{,\theta}\sin^{3} \theta
+\frac{1}{4}cU_{3}\sin^{3} \theta
-\frac{3}{16} (\gamma_{3,1}^{0})_{,\theta} \sin^{3} \theta
+\frac{3}{8}\gamma_{3,1}^{0}\cos \theta \sin^{2} \theta \nn
\end{eqnarray}
Since $G$ contains no $\log r$ terms, it follows that ${\cal Q}$ as defined in
(\ref{newlogconservedquantity}) is conserved. (We may again note that
if $\gamma_2 \equiv 0 \equiv \gamma_{3,1} $ this agrees with \cite{vdB}.)

\end{document}